\documentclass[aps,prd,twocolumn,showpacs,superscriptaddress,groupedaddress]{revtex4}  
\usepackage{graphicx}  
\usepackage{dcolumn}   
\usepackage{bm}        
\usepackage{amssymb}   
\usepackage{amsmath}
\usepackage{color}

\def\be{\begin{equation}}
\def\ee{\end{equation}}
\def\bea{\begin{eqnarray}}
\def\eea{\end{eqnarray}}

\begin{document}

\widetext
{\flushright{DESY 17-104}}


\title{A New Look at the $Y$ Tetraquarks and $\Omega_c$ Baryons in the Diquark Model}
\author{Ahmed~Ali}
\email{ahmed.ali@desy.de}
\affiliation{Deutsches Elektronen-Synchrotron DESY, D-22607 Hamburg, Germany}
\author{Luciano Maiani}
\email{Luciano.Maiani@cern.ch}
\affiliation{Theory Division, CERN, Geneva, Switzerland}
\author{Anatoly V. Borisov}
\email{borisov@phys.msu.ru}
\affiliation{Faculty of Physics, Moscow State University, 119991 Moscow, Russia}
\author{Ishtiaq Ahmed}
\email{ishtiaqmusab@gmail.com}
\affiliation{National Centre for Physics, Quaid-i-Azam University Campus, Islamabad, 45320, Pakistan}
\author{M. Jamil Aslam}
\email{muhammadjamil.aslam@gmail.com}
\affiliation{Department of Physics, Quaid-i-Azam University, Islamabad, 45320, Pakistan}
\author{Alexander Ya. Parkhomenko}
\email{parkh@uniyar.ac.ru}
\affiliation{Department of Theoretical Physics, P.\,G.~Demidov Yaroslavl State University, Sovietskaya 14, 150003 Yaroslavl, Russia}
\author{Antonio D. Polosa}
\email{antonio.polosa@roma1.infn.it}
\affiliation{Dipartimento di Fisica and INFN,  Sapienza  Universita di Roma, Piazzale Aldo Moro 2, I-00185 Roma, Italy}
\author{Abdur Rehman}
\email{Abdur.Rehman@fuw.edu.pl}
\affiliation{National Centre for Physics, Quaid-i-Azam University Campus, Islamabad, 45320, Pakistan}
\date{\today}

\begin{abstract}
We analyze the  hidden charm $P$-wave tetraquarks in the diquark model, using an effective Hamiltonian incorporating the dominant spin-spin, spin-orbit and tensor interactions. We compare with other $P$-wave system such as $P$-wave charmonia and the newly discovered $\Omega_c$ baryons, analysed recently in this framework. Given the  uncertain experimental situation on the $Y$ states, we allow for different spectra and discuss the related parameters in the diquark model.
In addition to the presently observed ones, we expect many more states in the supermultiplet of $L=1$ diquarkonia, whose
$J^{PC}$ quantum numbers and masses are worked out, using the parameters from the currently preferred  $Y$-states pattern.
The existence of these new resonances would be a decisive footprint of the underlying diquark dynamics.
\end{abstract}

\maketitle

\section{Introduction} 
\label{sec:introduction} 
 
The experimental discovery of exotic, hidden charm or beauty, states has opened a new field in  hadron spectroscopy.
The exotic  
states, called $X,Y,Z$ and~$P_c$,  have been analysed in a number of theoretical models. They have been claimed to be hybrid charmonia, molecules, disguised charmonium states, or just coupled-channel or  threshold effects, see \cite{Ali:2017jda, Esposito:2016noz,  Chen:2016qju, Guo:2017jvc,Lebed:2016hpi,Olsen:2017bmm} for recent reviews and references therein. 
We concentrate on the alternative diquark-antidiquark interpretation,  tetraquark in brief, introduced in~\cite{Maiani:2004vq,Ali:2011ug} following the light pentaquark  picture discussed in~\cite{Jaffe:2003sg}, which has the potential to include all exotic hadrons seen thus far in a single scheme.

The objects of our interest in this paper are the so called $Y$ states, with $J^{PC}=1^{--}$, described as tetraquarks with orbital angular momentum $L=1$ and  $L=3$.
$Y$-states have also been  interpreted as hadron molecules in~\cite{Wang:2013cya}, with $Y(4008)$ not foreseen in that case. The state $Y(4260)$ has also been advocated as an example of a $(c\bar{c})_8 g$ hybrid
\cite{Close:2005iz}. However,
evidence of two resonant structures in $e^+ e^- \to \pi^+ \pi^- h_c$ in the $Y(4260)$ region 
 has been presented by the BESIII collaboration~\cite{BESIII:2016adj}. 
 This would imply in the hybrid picture the existence of two states, nearby in mass,
  having  different $c\bar{c} $-spins.
~Another analysis of the $Y$ states in the diquark approach can be found in~\cite{Lebed:2016yvr}.

The univocal prediction of tetraquarks is that  here must be only four $Y$ states arising from the orbital angular momentum $L=1$ and no radial excitation, as shown in Table~\ref{ystates}. Parity requires $L$ to be odd, and charge conjugation requires the spin structure of these states to be symmetric under the exchange diquark $\leftrightarrow$ antidiquark, due to the factor $(-1)^L$  introduced by the exchange of coordinates. Besides those of Table~\ref{ystates}, there are  two other spin structures possible which are antisymmetric and in $P$ wave they would give $J^{PC}=1^{-+}$.  
One obtains $J^{PC}=1^{--}$  also from $L=3$, but this state is expected to be considerably heavier and we do not consider it.   
 \begin{table}
\begin{center}
\caption{$J^{PC}=1^{--}$ tetraquarks involving a diquark-antidiquark $\mathcal{Q}\bar{\mathcal{Q}}$ pair in $P$ wave.}
\vspace*{2mm}
{\begin{tabular}{cc}
\toprule
Label & 
$| S_{\mathcal{Q}}, S_{\bar{\mathcal{Q}}}; S, L \rangle_J$ \\
\colrule
\hline
$Y_1$ & $| 0, 0; 0, 1 \rangle_1$ 
\\
$Y_2$ & 
$\big( | 1, 0; 1, 1 \rangle_1 + | 0, 1; 1, 1 \rangle_1 \big)/\sqrt{2}$ 
\\
$Y_3$ & 
$| 1, 1; 0, 1 \rangle_1$ 
\\
$Y_4$ & 
$| 1, 1; 2, 1 \rangle_1$ 
\\
\hline\hline
\end{tabular}
}
\label{ystates}
\end{center}
\end{table}

In~\cite{Maiani:2014aja}, the four basic $L=1$ resonances with  $J^{PC}=1^{--}$ of the diquark-antidiquark spectrum were identified with $Y(4008)$, $Y(4260)$, $Y(4290)$ (a broad structure in the $h_c$ channel), or $Y(4220)$ (a narrow structure in the same channel) and a possible $\Lambda \bar \Lambda$ resonance around $4630$~MeV. The $Y(4360)$ and the $Y(4660)$, also known at that time, were tentatively considered to be $n=2$ radial excitations of $Y(4008)$ and $Y(4260)$, respectively. 
~Since that paper appeared, the experimental situation has evolved considerably. 
The status of the $Y(4008)$ is no longer established,  and the $Y(4260)$ is now claimed by BESIII as a double humped structure \cite{Ablikim:2016qzw}, which is resolved into two resonances: a lower component, $Y(4220)$, with  observed decays into $h_c~ \pi^+ \pi^-$ and $\chi_{c0}~\omega$, and a higher component, $Y(4330)$, which decays into $J/\psi ~\pi^+ \pi^-$.  
On the other hand, it was also observed that $Y(4630)$ and $Y(4660)$ could be fitted as a unique resonance, mainly decaying into $\Lambda\bar \Lambda$~\cite{Cotugno:2009ys}.

  In conclusion, there seem to be at present two  favoured scenarios, SI and SII, both comprising of four $Y$ states and based essentially on the Belle, BaBar and BESIII data, namely:
\begin{itemize}
\item {\bf Scenario~I}:\quad $Y(4008),Y(4260), Y(4360),Y(4660)$,   favoured in~\cite{Ali:2017jda}; 
\item {\bf Scenario~II}:\quad $Y(4220), Y(4330), Y(4390), Y(4660)$, favoured in~\cite{Olsen:2017bmm}.
\end{itemize}

SI assumes $Y(4008)$ to be a real resonance and it features $Y(4260)$ as a single state. Belle~\cite{Abe:2006hf} finds that data are better fit with two resonances, $Y(4260)$ and $Y(4008)$, and the width of the $Y(4008)$ is found to be a factor 2 larger than that of the $Y(4260)$. 
However, the $Y(4008)$ has been seen so far by Belle only and current analysis of this resonance from BESIII is inconclusive~\cite{Ablikim:2016qzw}. 
On this basis, SII discards the $Y(4008)$ and it features the two lines resolving the $Y(4260)$, according to BESIII~\cite{Ablikim:2016qzw},  $Y(4220)$ and $Y(4330)$, as lowest resonances.  
The $Y(4360)$ and $Y(4390)$ appearing in SI and SII respectively,   are 
considered as the same resonance seen in different experiments. Similarly, in both SI and SII, one considers $Y(4660)$ and the proposed $\Lambda-\bar \Lambda$ resonance at $4630$~MeV to correspond to the same state.

 The spectra of the two scenarios extend over $400-600$ MeV, and one could wonder if this is consistent with these states belonging to a single, fine structure multiplet, given that $P$-wave charmonium states are all within an interval of about $100$ MeV.  However, the difference can be defended  by the different composition of the tetraquarks w.r.t. the charmonia, as explained below.

First, in tetraquarks the total quark spin goes up to $S=2$, which amplifies the range of the spin-orbit and tensor couplings, and the effect of the tensor force in tetraquarks has not been investigated so far. Secondly, the constituents of $Y$ states are diquarks and antidiquarks with spin $0$ and $1$, the "good" and "bad" diquarks in Jaffe's terminology, see~\cite{Jaffe:2003sg}. For $S$-wave tetraquarks,  the mass difference between $Z(4020)$ and $Z(3900)$ results in a mass difference of "bad"to "good" $[cq]$ diquarks of about $120$~MeV, and the $Y$ states contain from zero to two "bad" diquarks. This splits the
masses of the different components of a tetraquark multiplet considerably more than in the conventional charmonia.
Note that, the diquark mass difference is related to the spin-spin coupling between the charm and the light quark in the diquark, which comes out to be $3-4$ times the $c-q$ coupling in the charmed baryons ~\cite{Maiani:2014aja}. In QCD, these couplings are proportional to the quark overlap probability, $|\psi(0)|^2$, and the result is simply understood to indicate a closer packing of diquarks in the tetraquarks than in baryons. 
 
The upper range of masses in the two scenarios goes into the region where radial excitations of the lowest $P$-wave tetraquarks are expected and one may wonder if the highest and, possibly, the next to highest $Y$ states may be the radial excitations of the two lowest ones.  For definiteness we assume this not to be the case. 

It is still possible, however, that a better experimental resolution may substantiate the differences observed between the $4360-4390$ and the $4630-4660$ peaks so as to indicate the presence of one or more radial excitation in the region, as was assumed in~\cite{Maiani:2014aja}.  
We examine the issue of possible radial excitations in Section~\ref{variations}.  A similar issue has been raised for the excited $\Omega_c =css$ states,
whose mass spectrum has been measured by  the LHCb collaboration \cite{Aaij:2017nav}, and  confirmed recently by 
  Belle \cite{Yelton:2017qxg}, except perhaps the $\Omega_c(3119)$. 
The LHCb mass spectrum is
  discussed in a number of papers \cite{Wang:2017vnc,Padmanath:2017lng,Aliev:2017led,Karliner:2017kfm}, following the analysis of ~\cite{Karliner:2017kfm}, in which all  observed five states are assumed to be $P$-wave $c$ quark and $ss$ diquark. Also here,  the highest mass states overlap with the $2S$, positive parity, radial excitations of the $S$-wave $\Omega_c$~\cite{Ebert:2011kk,Agaev:2017jyt}.  This issue will
be clarified as and when the $J^P$ quantum numbers of the excited $\Omega_c$ states are experimentally determined.

Within the two Scenarios given above, we work out the mass  spectrum derived from the spin-orbit, spin-spin and tensor coupling interactions, the latter was not included in~\cite{Maiani:2014aja}. The principal aim is to investigate whether the tetraquark picture may provide a satisfactory description of the presently determined $Y$-states in the $c\bar{c}$ sector, eventually distinguishing between SI and SII.
For instance, since parameters are obtained from the solution of a second order equation, the mass formulae could produce complex parameters.  In the case that there is no real-parameter solution possible, the underlying theoretical framework, namely the diquark picture, can be excluded as a physical template for the Y states. We use the reality condition on the parameters to
eliminate some alternative assignments, as discussed below.

On the positive side, one would expect the value of the chromomagnetic, spin-spin coupling inside the diquark, $[\kappa_{cq}]_P$, to be close to the analogous parameter derived for the S-wave tetraquarks, which is $[\kappa_{cq}]_S \simeq 67$ MeV as discussed in~\cite{Maiani:2014aja}. For tight diquarks, this parameter should not be too much affected by the addition of one unit of orbital angular momentum. 
In addition, comparison with the $S$-wave tetraquark masses can give the energy for the excitation of one unit of orbital momentum. 

The diquark mass being very similar to the charm quark mass, one expects the excitation energy of the tightly bound Y states to be similar to the one obtained from the comparison of  the $P$ and $S$-wave charmonia and for the $P$-wave $\Omega_c$ states. This expectation is indeed satisfied by one solution in each of the two scenarios. 

The tetraquark scheme predicts several other negative parity states with different $J$, and another $1^{--}$ resonance arising from $L=3$. Ignoring the $L=3$ state, which is presumably rather heavy, we comment in the end on the composition of the full $L=1$ supermultiplet and give an estimate of the expected masses.

The plan of the paper is as follows. 
In Sect.~\ref{omegaham}, we repeat the analysis of the five $L=1$ charmed-baryons~$ \Omega_c$, whose mass spectrum has recently been measured by  the LHCb collaboration \cite{Aaij:2017nav}, 
following the analysis of ~\cite{Karliner:2017kfm}.
In this connection, we offer an alternative calculation of the
tensor couplings in terms of the Wigner's 6-$j$ symbols. 

In Sect.~\ref{tetraham}, we introduce the two scenarios compatible with the present data, derive the mass formulae for the $Y$ states and obtain the parameters of the Hamiltonian from the mass spectra. The role of radial excitations is discussed in Sect.~\ref{variations}. Error analysis and correlations among the parameters are presented
in Sect.~\ref{errors}. 
Results are discussed in Sect.~\ref{discussion}.

A picture of the full $L=1$ multiplet is reported in Sect.~\ref{fullmulti} and Conclusions are presented
 in Sect.~\ref{conclu}.

In Appendix~\ref{6jsymbols} we derive  the tensor couplings for the $\Omega_c$ baryons and diquarkonium, using Wigner's 6-$j$ symbols. Correlation matrices in the analysis of the data on the $Y$ states are given in Appendix~\ref{app:correlation}.
 
\section{Effective Hamiltonian for the $\Omega_c$~baryons}
\label{omegaham}

In the diquark-quark description, the Hamiltonian for the $\Omega_{c}$ states can be written as~\cite{Karliner:2017kfm}:
\begin{eqnarray}
H_{\rm eff}&=&m_{c} + 2 m_{s}+2\kappa_{ss}{\bm{S}_{s_1}\cdot \bm{S}_{s_2}}+ \frac{B_{\mathcal Q}}{2}{\bm{L}^2}+ V_{SD}, \label{Hamiltonian-OC} \\
V_{SD} &=& a_1 {\bm{L}\cdot \bm{S}_{[ss]}} + a_2  \bm{L}\cdot \bm{S}_c + b~\frac{  S_{12}}{4}+ c\, {\bm {S}_{[ss]}\cdot \bm{S}_c}. \nonumber
\end{eqnarray}
In Eq. (\ref{Hamiltonian-OC}), $m_{c}$ and $m_{s}$ are the masses of the $c$ and the $s$ quarks, respectively,
 $\kappa_{ss}$ is the spin-spin coupling of the quarks in the diquark, and ${\bm{L}} $ is the orbital angular
 momentum of the diquark-quark system. 
  The coefficients $a_1$ and $a_2$ are the strengths of
 the spin-orbit terms involving the spin of the diquark $ \bm{S}_{[ss]}$ and the charm-quark spin $ \bm{S}_c $, respectively,
 $c$ is the strength of the spin-spin interaction between the diquark and the charm quark, and
 $ S_{12}/4 $ represents the tensor interaction, defined by 
 \begin{equation}
\frac{S_{12}}{4}=Q({\bm   S}_1,   {\bm   S}_2)=3({\bm S}_1\cdot {\bm n})({\bm S}_2\cdot {\bm n})-({\bm S}_1\cdot {\bm S_2}), \label{tensor}
\end{equation}
where $\bm S_{1}$ and $\bm S_{2}$ are the spins of the diquark and the charm quark, respectively, and ${\bm n}={\bm r}/r$ is the unit vector along the radius vector of a particle. 
 
The  scalar  operator of Eq. (\ref{tensor}) can  be expressed as the product
$3S_{1}^{i}S_{2}^{j}N_{ij}$, where the tensor operator is
\begin{equation}
N_{ij} = n_{i} n_{j} -\frac{1}{3}\delta_{ij}.
\end{equation}
To compute the  matrix elements of this operator between states with the same fixed value $L=1$  one uses the identity from Landau and Lifshitz  \cite{Landau:1977} (see also \cite{Ebert:2002ig}):
 \begin{eqnarray}
&&\langle   N_{ij}   \rangle   =   a(L) (  L_i  L_j+L_j  L_i
- \frac{2}{3}\delta_{ij} L(L+1));\nonumber \\
&&a(L)=\frac{-1}{(2L-1)(2L+3)}, \label{Landaulif}
 \end{eqnarray}

One finds
 \begin{eqnarray}
 &&\langle \frac{S_{12}}{4} \rangle_{(L=1)}=3 ~a(L) \nonumber \\
 &&\times \left[({\bm L \cdot}{\bm  S}_1)({\bm L \cdot}{\bm S}_2)+({\bm L \cdot}{\bm S}_2)({\bm L \cdot }{\bm S}_1)-\frac{4}{3}({\bm S}_1{\bm \cdot}{\bm S}_2)\right],\nonumber \\
 \label{master3}
 \end{eqnarray}
which requires the matrix elements
\begin{equation}
\langle S^\prime, L; J|{\bm L\cdot }{\bm S}_{1,2} |S, L; J\rangle.\label{matrel}
\end{equation}

The latter can be computed by applying the operators ${\bm L\cdot}{ \bm S}_{1,2}$ to the products of three angular momentum states, see~\cite{Karliner:2017kfm}. More effectively, one can use Wigner's 6-$j$ symbols (now implemented  in computer codes), as is customary for analogous cases in atomic and nuclear physics and is explained in Appendix~\ref{6jsymbols}. 

 In either way, one obtains: 
\begin{eqnarray}
&&J=1/2:\quad \frac{1}{4}\, \langle S_{12} \rangle=\left(\begin{array}{cc} 0 & \frac{1}{\sqrt{2}}\\  \frac{1}{\sqrt{2}} & -1\end{array}\right),\nonumber \\
&&J=3/2:\quad  \frac{1}{4}\, \langle S_{12} \rangle=\left(\begin{array}{cc} 0 &- \frac{1}{2\sqrt{5}}\\  -\frac{1}{2\sqrt{5}} & \frac{4}{5} \end{array}\right), \label{half-threehalf}\\
&&J=5/2:\quad  \frac{1}{4}\, \langle S_{12} \rangle= -\frac{1}{5} .\notag
\end{eqnarray}

After diagonalisation, we get the mass corrections arising from the Hamiltonian~(\ref{Hamiltonian-OC}). We remind that  in all the five states, there is the common mass term
\be
  M_{0} \equiv m_{c} + 2 m_{s} + \frac{1}{2}\kappa_{ss} + B_{\mathcal Q} \label{eq:M0}.
\ee
In order to determine the parameters $a_{1},\; a_{2},\; b$ and~$c$, Karliner and Rosner~\cite{Karliner:2017kfm} have used the spin averaged mass  and have worked with the mass differences of the five $\Omega_{c}$ states.
We reproduce their values,  summarised in Table \ref{omega-table}, where we have also given the value of $M_{0}$.
 \begin{table}
 \begin{center}
\caption{Values of the parameters $a_1, a_2$, $b$, $c$ and $M_{0}$ (in MeV), determined from the masses 
of the $\Omega_c$ baryons given in~\cite{Aaij:2017nav} and the spin assignments from~\cite{Karliner:2017kfm}.}
\label{omega-table}
{\normalsize
{\begin{tabular}{|c|c|c|c|c|}
\hline
$a_{1}$ & $a_2$  &  $b$ & $c $ & $M_{0}$ \\
\hline
$26.95$ & $25.75$  &  $13.52$ & $4.07$ & $ 3079.94$  \\ 
\hline%
\end{tabular}
}
}
\end{center}

\end{table}

\section{Effective Hamiltonian for $Y$~Tetraquarks}
\label{tetraham}  

$Y$-states have the quark content $[cq]_{\bar{3}}[\bar{c}\bar{q}]_{3}$, where the subscripts denote color representations.
Tetraquarks with $J^{PC}=1^{--}$ are obtained for $L=1,\,3$. Spin wave functions are given in Table~\ref{ystates}, in the basis ${\bm S}_{\mathcal Q}={\bm S}_{[cq]},~{\bm S}_{\bar {\mathcal Q}}={\bm S}_{[\bar{c}\bar{q}]},~{\bm L}$ with ${\bm S}= {\bm S}_{\mathcal Q}+{\bm S}_{\bar {\mathcal Q}}$ and ${\bm J}={\bm S}+{\bm L}$.

We extend the Hamiltonian of $P$-wave tetraquarks given in \cite{Maiani:2014aja} by including the tensor coupling contribution
\begin{eqnarray}
H_{\rm eff}&=&2m_{\mathcal Q}
+ \frac{B_{\mathcal Q}}{2}  \bm{L}^2 -3\kappa_{cq}
+2a_Y  \bm{L}\cdot \bm{S}  +b_Y\frac{ S_{12}}{4}  \nonumber\\
&+& \kappa_{cq} \big[2( \bm{S_{q}}\cdot \bm{S_{c}} 
 +  \bm{S_{\bar q}}\cdot \bm{S_{\bar c}} ) +3 \big],  \label{Hamiltonian}
\end{eqnarray}
 $S_{12}$ is defined as in Eq.~\eqref{tensor}, with $\bm S_{1,2}$ representing the spins of the diquark and antidiquark, and
 \be
 2( \bm{S_{q}}\cdot \bm{S_{c}}+  \bm{S_{\bar q}}\cdot \bm{S_{\bar c}})+3=2N_1
 \label{baddqn}
 \ee
 where $N_1$ is the number of spin $1$, "bad", diquarks.
 Comparing to~\eqref{Hamiltonian-OC}, we see that 
in this case the coefficients~$a_{1}$ and~$a_{2}$ are $a_1 = a_2 \equiv 2 a_Y$ 
due to charge conjugation invariance. 
 The spin-spin interaction
between diquark and antidiquark is neglected here since in $P$-wave the overlap probability is suppressed~\cite{Maiani:2014aja}. In the $\Omega_c$ case, the spin-spin interaction, represented by~$c$, Table~\ref{omega-table}, is similarly suppressed and the same happens in $P$-wave charmonia.

The calculation of the matrix elements of the $\bm L\cdot \bm S_X$ operator, with
 $\bm S_X=\bm S_{[cq]}$ and $\bm S_{[\bar c\bar q]}$, is described in Appendix A, Eqs.~\eqref{mat4q1} and~\eqref{mat4q2}. We note here that: 
\begin{itemize}
\item tensor couplings are non vanishing only for the states with ${ S}_{\mathcal Q}={ S}_{\bar {\mathcal Q}}=1$;
\item the operator ${\bm L}\cdot{\bm S}_{\mathcal Q}$ is not invariant under charge conjugation and it does mix the states $Y_3$ and $Y_4$, with a $J^{PC}=1^{-+}$ state with the composition:
\be
Y^{(+)} = |1,1;1,1\rangle_1.
\label{cplus}
\ee
\item $Y^{(+)}$ appears as an intermediate state in the products $(\bm L\cdot \bm S_{\mathcal Q})(\bm L\cdot \bm S_{\mathcal{\bar  Q}})$ and $(\bm L\cdot \bm S_{\mathcal {\bar Q}})(\bm L\cdot \bm S_{\mathcal Q})$, giving contribution to both diagonal and non diagonal terms;  charge conjugation invariance is of course restored when making the sum of the two products, which is block diagonal in the basis $(Y_3,~Y_4)$ and $Y^{(+)}$.
\end{itemize}
In conclusion, we have to consider the full $(3\times 3)$ matrix $\bm L\cdot \bm S_{[cq]}$. Using~\eqref{6jcalculation} and~\eqref{assdq} we find:
\begin{eqnarray}
&&\langle \bm L\cdot \bm S_{[cq]}\rangle_{J=1}=\langle 1,S^\prime; 1|  \bm L\cdot \bm S_{[cq]}|1,S; 1\rangle \notag \\
&& = \sqrt{(2S+1)(2S^\prime+1)} \sum_{j=0}^2 \left ( 2 j + 1 \right ) \notag \\
&& \times\frac{1}{2}\left [ j(j+1) - 4 \right ]~ \begin{Bmatrix}1& 1& j\\1&1&S^\prime\end{Bmatrix} \begin{Bmatrix}1& 1& j\\1&1&S\end{Bmatrix}, \label{form6j} 
\end{eqnarray}
where $S,~S^\prime=0,~1,~2$ and the curly brackets denote Wigner's 6-$j$ symbols~\cite{edmonds}.
After Eqs.~(\ref{master3}) and (\ref{Hamiltonian}), tensor couplings over the $Y_3-Y_4$ states are represented by the non-diagonal matrix:
\be
\frac{ \langle S_{12}\rangle}{4} =\left( 
\begin{array}{cc} 
 0 & 2/\sqrt{5} \\ 
 2/\sqrt{5} & -7/5%
\end{array}%
\right).  \label{splitting-matrix}
\ee

The eigenvalues of the mass matrix of $Y$-states derived from Eqs.~(\ref{Hamiltonian}) and~(\ref{splitting-matrix}), are written as: 
\begin{eqnarray}
M_1=M(Y_1) &=& M_{00} -3\kappa_{cq} \equiv \widetilde{M}_{00},\nonumber\\
M_2=M(Y_2)&=& \widetilde{M}_{00} - 2a_Y + 2 \kappa_{cq},\label{Y-states}\\
M_3&=& \widetilde{M}_{00} + 4 \kappa_{cq} + E_{+},\nonumber\\
M_4 &=& \widetilde{M}_{00} + 4\kappa_{cq} + E_{-}. \nonumber
\label{massform1}
\end{eqnarray}
and $M_{00} = 2 m_{\mathcal{Q}}+ B_{\mathcal Q} $.
We have made explicit that the states $Y_{1,2}$ in Table~\ref{ystates} are eigenstates of the mass matrix, while $M_{3,4}$ are the eigenvalues of the matrix
\be
2a_Y\langle {\bm L}\cdot {\bm S}\rangle +b_Y\langle S_{12}\rangle/4 ,
\ee
with
\begin{eqnarray}
&&E_{\pm} = \frac{1}{10}\nonumber \\
&&\times  \left(-30 a_Y - 7 b_Y\mp \sqrt{3} \sqrt{300 a_Y^2 + 140 a_Yb_Y + 43 b_Y^2}\right), \nonumber\\
&&M_3 + M_4 = 2(\widetilde{M}_{00} + 4 \kappa_{cq}) + \frac{1}{5} (-30 a_Y - 7 b_Y)\nonumber\\
&&\hspace*{11mm} = 2(\widetilde{M}_{00} + 4 \kappa_{cq}) + E_{+} + E_{-}, \nonumber\\
&& M_4 - M_3 =  \frac{\sqrt{3}}{5} \sqrt{300 a_Y^2 + 140 a_Yb_Y + 43 b_Y^2}\nonumber\\
&&\hspace*{11mm} = E_{-}-E_{+} \geq 0.
\label{Y34-states}
\end{eqnarray}

In the scenario SI,  
we take the four $J^{PC}=1^{--}$ $Y$ states to be $Y(4008)$, $Y(4260)$, $Y(4360)$ and $Y(4660)$, with masses (all in MeV)
\begin{eqnarray}
M_1 &=& 4008 \pm 40^{+114}_{-28}, \quad M_2 = 4230 \pm 8, \nonumber\\
M_3 &=& 4341 \pm 8 ,\quad\quad\quad M_4 = 4643 \pm 9. \label{Ymasses}
\end{eqnarray}
Masses are taken from PDG~\cite{Olive:2016xmw}, except for the  $Y(4008)$, which is from Belle \cite{Abe:2006hf}. 

In the scenario SII~%
the masses of $Y(4220),\; Y(4330)$, $Y(4390)$ and  $Y(4660)$ are (all in MeV):
\begin{eqnarray}
M_1 &=& 4219.6 \pm3.3\pm 5.1, \quad M_2 = 4333.2 \pm 19.9 ,\nonumber\\
M_3 &=&  4391.5 \pm 6.3, \quad M_4 = 4643 \pm 9, \label{Ymasses-used}
\end{eqnarray} 
{\it i.e.} the state 
with the mass~$M_4$ is the same as in SI.

 For $S$-waves~\cite{Maiani:2014aja}, the spin-spin interaction gives a larger mass to $S=1$ diquarks with respect to $S=0$ ones and the same for antidiquarks. For this reason, it is natural to associate $Y_1$ and $Y_2$ with the two lightest particles in increasing mass order,  $M_1<M_2$. This association is forced by the fact that if we exchange the role of $Y_1$ and $Y_2$, we obtain only complex solutions for the parameters of the Hamiltonian.

  In the case of SII, the association agrees with the fact that  $Y_1$ has a sizeable component over the state with vanishing $c-\bar c$ spin, in agreement with  the observed decays of $Y(4220)$ into $h_c$, while $Y_2$ has pure $c$-$\bar c$ spin equal one, in line with the observed decays of $Y(4330)$ into $J/\psi$. The assignment allows also to describe the decay $Y(4330)\to X(3872)+\gamma$ as an allowed electric-dipole transition, given that $Y_2$ has the same spin structure 
  as one attributes in the model to $X(3872)$. 

On the other hand, for $Y_3$ and $Y_4$, both containing two spin $1$ diquarks, we shall allow both possibilities: $Y_3\leftrightarrow M_3;~Y_4\leftrightarrow M_4$, with $M_3 < M_4$, or the other way round. 

\begin{table}
 \begin{center}
\caption{Values of the parameters in the scenarios I (SI) and II (SII) and $\pm 1 \sigma$ errors  (all in MeV). Here,~$c1$ and~$c2$ refer to the two solutions of the secular equation for $Y_{3,4}$}.
{\normalsize
{\begin{tabular}{|c|c|c|c|c|}
\hline
 & $a_Y$  &  $b_Y$ & $\kappa_{cq}$ & $M_{00}$   \\ 
\hline%
SI (c1)  & $-22 \pm 32$ &  $-89 \pm 77$ &  $89 \pm 11$  & $4275 \pm 54$ \\ 
\hline%
SI (c2)  &  $48 \pm 23$ &   $11 \pm 91$ & $159 \pm 20$  & $4484 \pm 26$ \\ 
\hline%
SII (c1) &  $-3 \pm 18$ & $-105 \pm 32$ &   $54 \pm 8$  & $4380 \pm 25$ \\ 
\hline%
SII (c2) &   $48 \pm 8$ &  $-32 \pm 47$ &   $105 \pm 4$ & $4535 \pm 10$ \\ 
\hline%
\end{tabular}
}
}
\label{ali:tbl3}
\end{center}
\end{table}

Before proceeding to the estimate of the values of the parameters $M_{00}$, $a_Y$, $\kappa_{cq}$ and~$b_Y$, we first note  their possible interdependence on each other. 
From Eq.~(\ref{Y34-states}) for $M_4 - M_3$ follows that
this mass difference is invariant under the simultaneous sign change $(a_Y, b_Y) \to (-a_Y, -b_Y)$. Hence, from this
mass difference alone, we have two solutions:  $a_Y <0$ and $a_Y >0$. We shall call them case 1 and case 2, respectively.
In line with the analysis for the  $\Omega_{c}$ states, given in Table \ref{omega-table}, only $a_Y >0$ should be kept.
This is also the choice suggested by the natural mass ordering, in which the $J=3$ state
 should have a higher mass than the $J=1$ states.
So, the only physically acceptable solution is the one which has positive value of $a_{Y}$ irrespective of the sign of the value of $b_{Y}$. 
 
However, as the errors on some of the masses are large,
we shall see below that, including the errors, solutions whose central values have  $a_Y <0$, are also allowed.  
\noindent
In addition to Eq.~(\ref{Y34-states}), the mass difference $M_{2} - M_{1}$ provides a constraint 
on the parameters $a_{Y}$ and $\kappa_{cq}$:
\begin{equation}
M_2 - M_1 =  2(\kappa_{cq} - a_Y).\label{Y12-states}
\end{equation}
Thus, in both the scenarios for the $Y_i$ masses, $\kappa_{cq} > a_Y$, with the two approaching each other as this
mass difference decreases.

The central values of the parameters $a_Y,~b_Y,~\kappa_{cq}$, and $M_{00}$  are determined from the masses given in Eq.~(\ref{Ymasses}) for SI and in Eq.~(\ref{Ymasses-used}) for SII and presented in Table~\ref{ali:tbl3}.  In each scenario, we indicate with (c1) and (c2) the two solutions obtained from the secular equation for $Y_3$ and $Y_4$.

\section{Variations on the theme}\label{variations}

We briefly comment on radial excitations, considering a Scenario III, proposed in~\cite{Lebed:2016hpi}, which 
envisages the confirmation of $Y(4008)$ and the doubling of $Y(4260)$:
\begin{itemize}
\item {\bf Scenario~III} ~\cite{Lebed:2016hpi}:~$Y(4008),Y(4230),Y(4330),Y(4390)$, $Y(4660)$.
 \end{itemize}
 
Given the masses of the first three states, we obtain the parameters of the Hamiltonian as functions of the mass of the fourth ground state, $M_4$. The parameters are real if this mass is such as to make positive the radicand in Eq.~(\ref{Y34-states}). Numerically, this implies:
\be
M_4 \geq 4450~{\rm MeV}.
\label{reality}
\ee 
This is consistent with the fourth ground state being $Y(4660)$, with the parameters very similar to those of Scenario I, but not with $Y(4390)$, which has to be the radial excitation of  $Y(4008)$.

Yet another possibility is to take the first three states of  the Scenario II and leave undetermined $M_4$. The reality condition
 gives a result close to (\ref{reality}), however with a parameter $\kappa_{cq}$ 
 a bit  smaller than expected from the $S$-wave masses. Assuming the fourth state to be the $Y(4660)$, we go 
back to the Scenario II with $\kappa_{cq} \simeq 54$~MeV, which is in the acceptable range.
In both alternatives considered, the full range of masses in Scenarios I and II is acceptable for $P$-wave ground states.

\begin{figure*}[tb] 
\begin{center}
\includegraphics[width=0.37\textwidth]{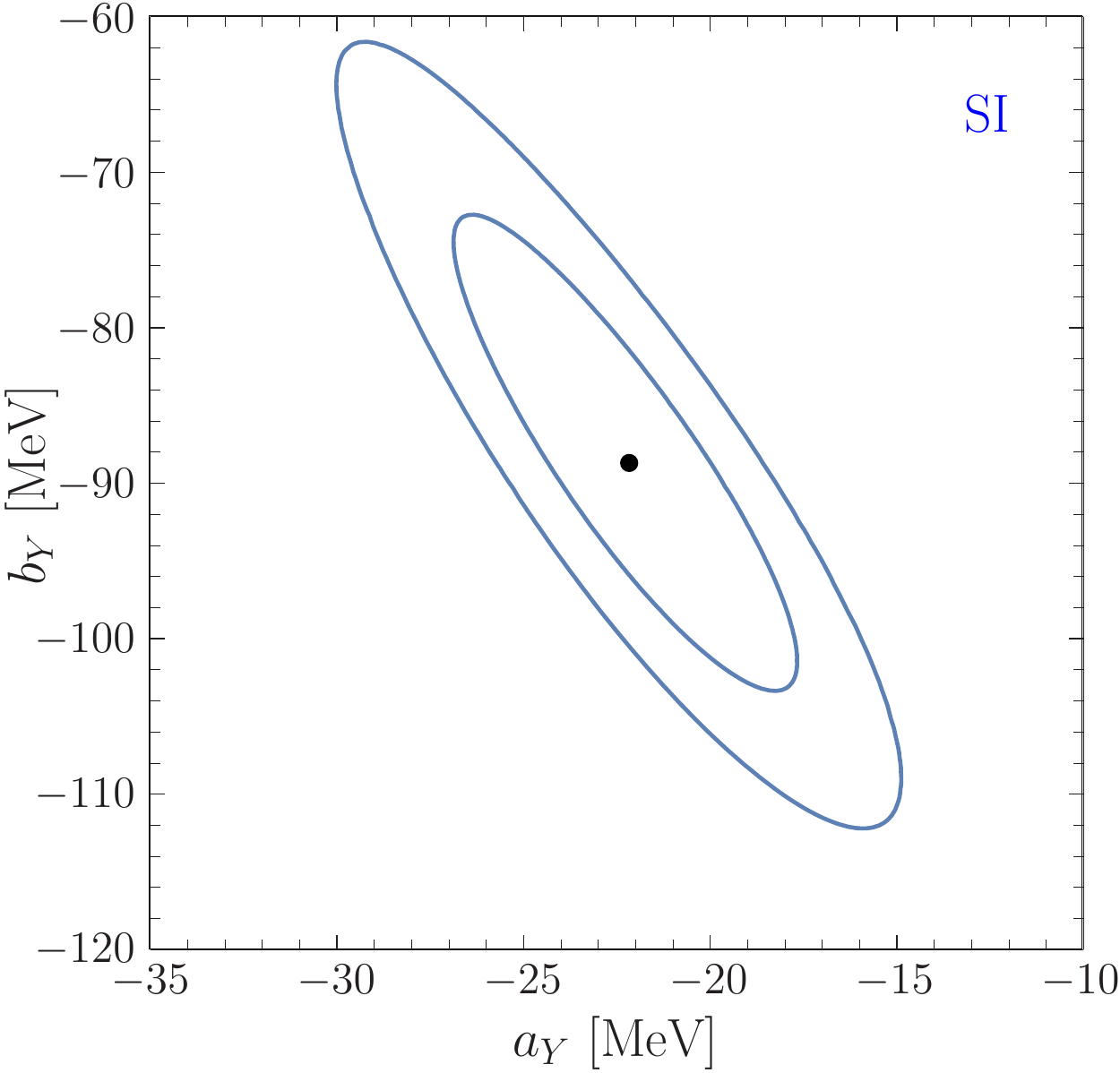} 
\quad 
\includegraphics[width=0.37\textwidth]{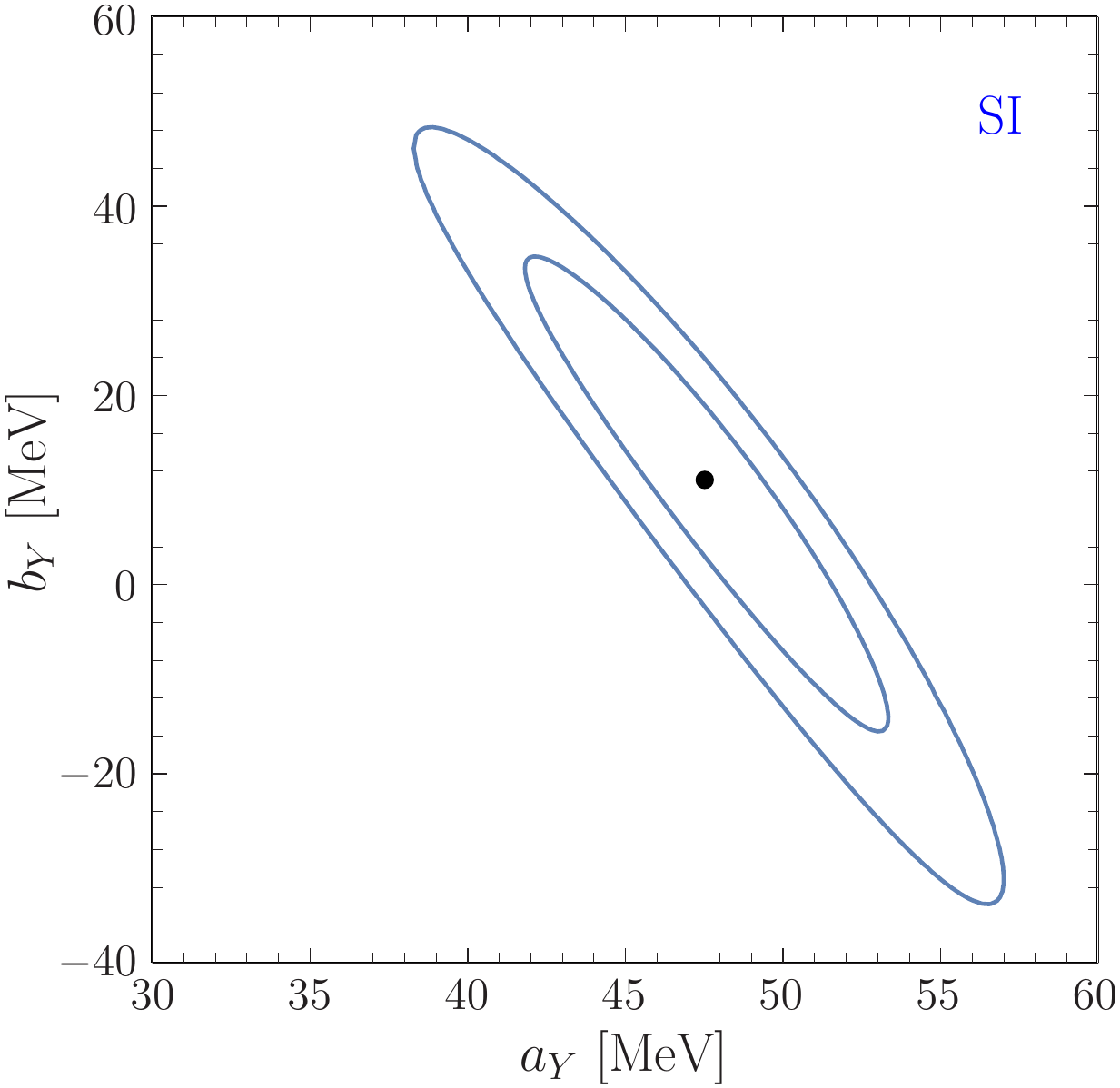} \\ 
\includegraphics[width=0.37\textwidth]{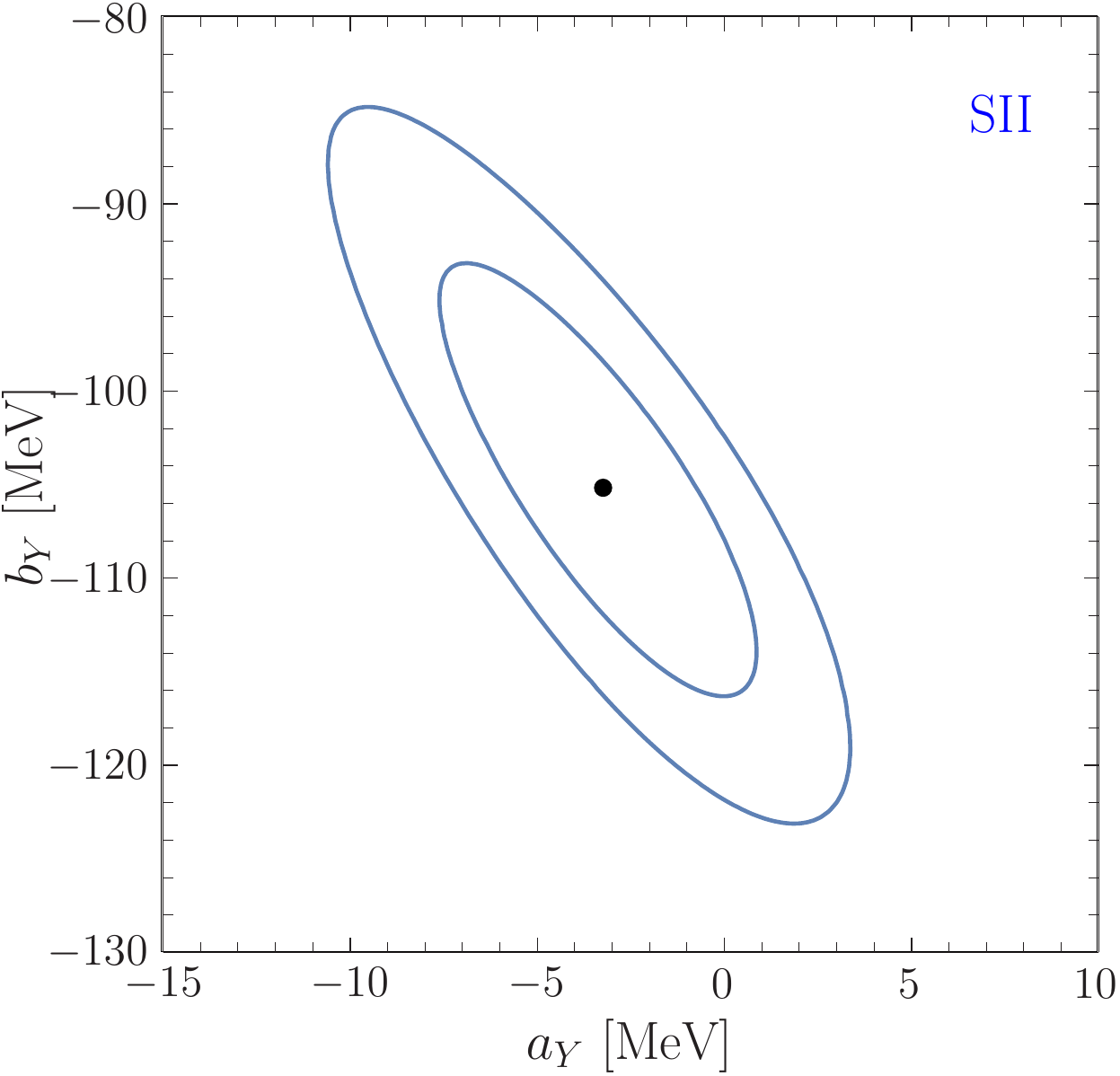} 
\quad 
\includegraphics[width=0.37\textwidth]{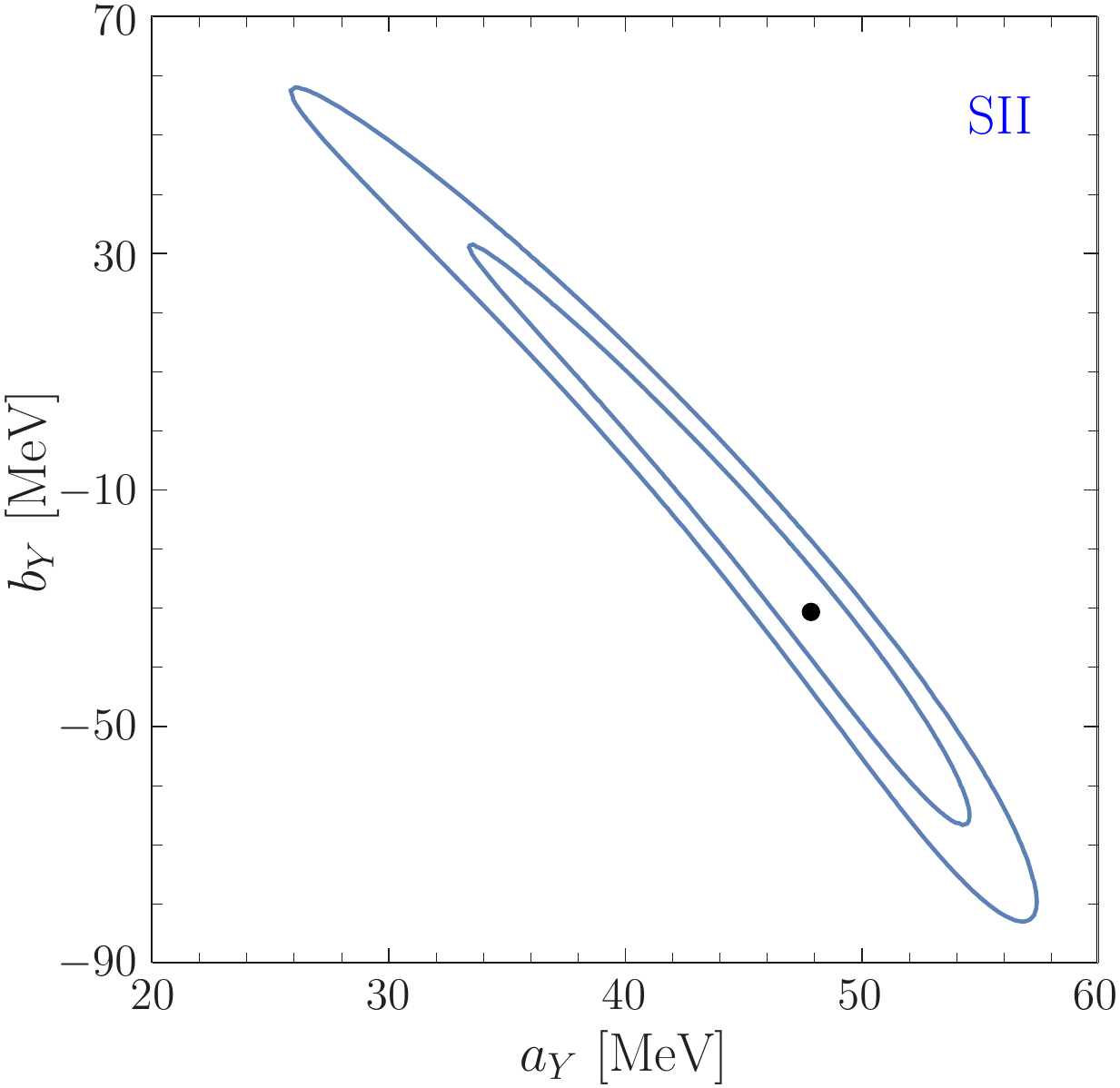} 
\caption{
$1\sigma$- and $2\sigma$-contours in the $a_Y - b_Y$ parameter plane 
corresponding to~68.3\% and~95.4\%~C.L. for the scenario~I (SI) 
in the top two frames and the scenario~II (SII) in the bottom two frames.
The dot in each frame shows the position of the best-fit value which 
is the minimum of the $\chi^2$-function. The best-fit value of~$a_Y$ 
is negative (case~1) in the left panels and positive (case~2) in the 
right panels. 
}  
\label{fig:correlations-aY-bY}
\end{center}
\end{figure*}

\begin{figure*}[tb] 
\begin{center}
\includegraphics[width=0.37\textwidth]{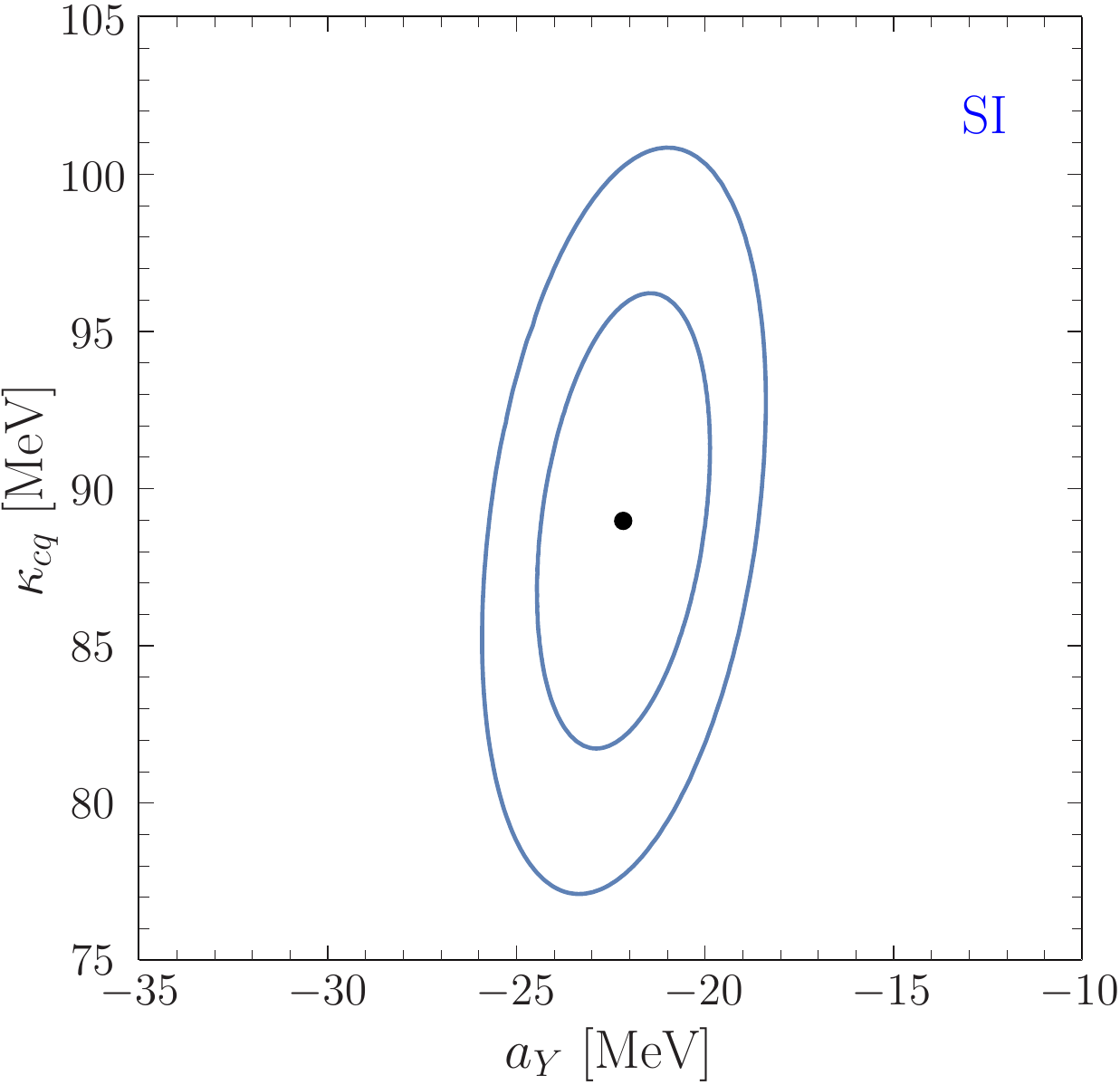} 
\quad 
\includegraphics[width=0.37\textwidth]{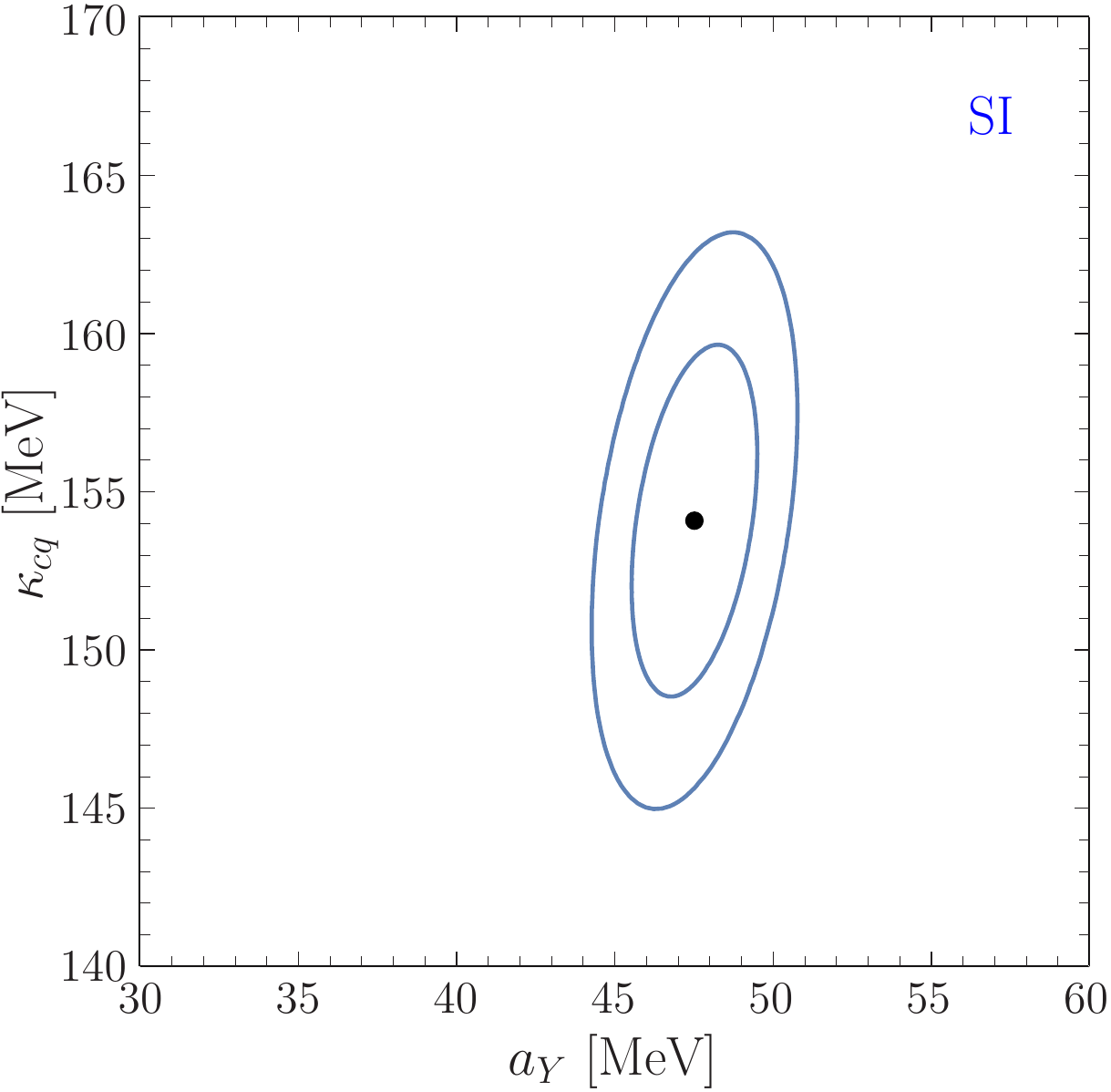} \\ 
\includegraphics[width=0.37\textwidth]{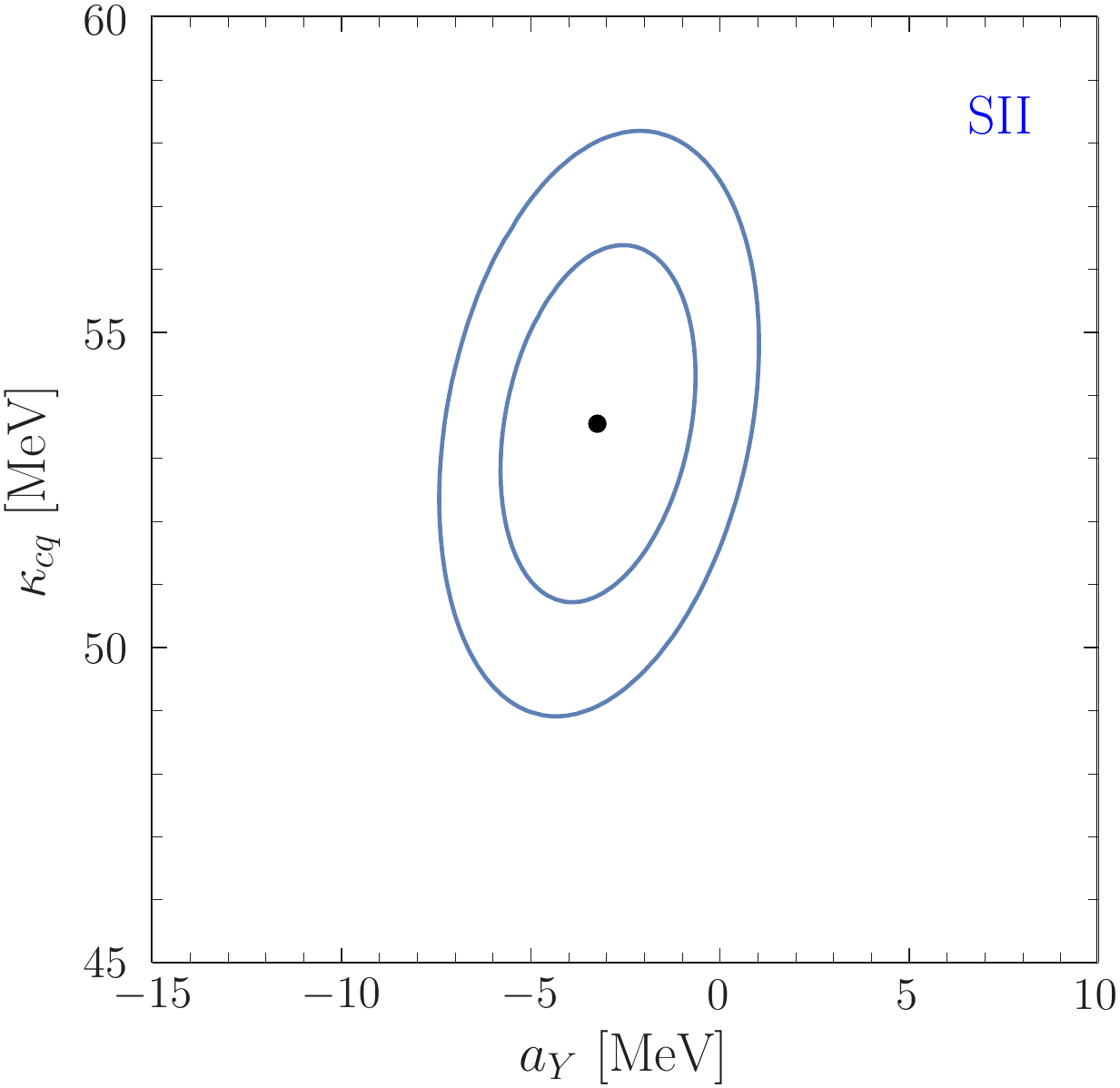} 
\quad 
\includegraphics[width=0.37\textwidth]{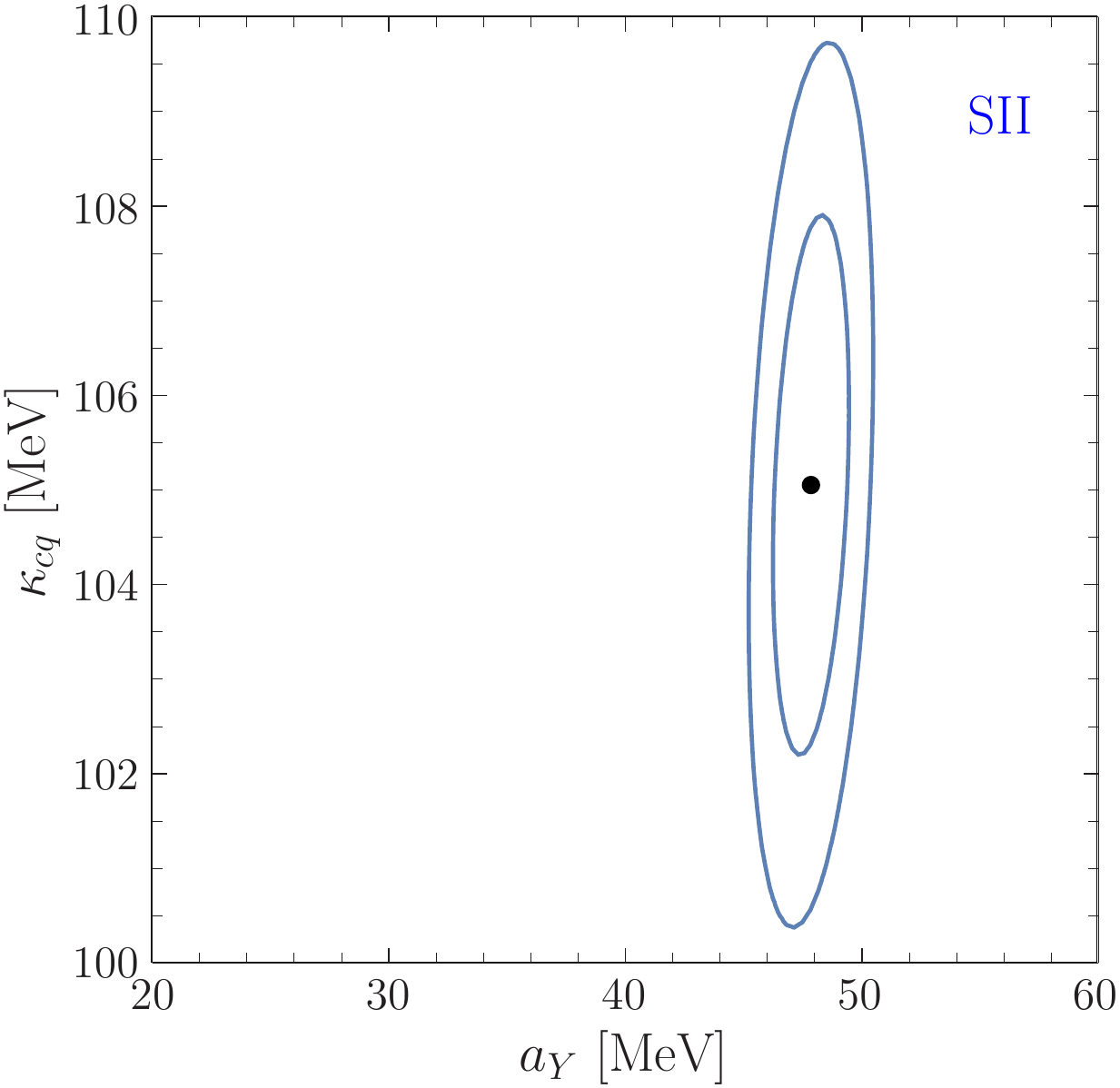} 
\caption{
$1\sigma$- and $2\sigma$-contours in the $a_Y - \kappa_{cq}$ parameter 
plane corresponding to~68.3\% and~95.4\%~C.L. for the scenario~I (SI) 
in the top two frames and the scenario~II (SII) in the bottom two frames.
The dot in each frame shows the position of the best-fit value which 
is the minimum of the $\chi^2$-function. The best-fit value of~$a_Y$ 
is negative (case~1) in the left panels and positive (case~2) in the 
right panels. 
}  
\label{fig:correlations-aY-kcq}
\end{center}
\end{figure*}

\section{Errors and Correlations}\label{errors}
To work out errors and correlations among the parameters, we have used the method 
of least squares to determine the best-fit values and the covariance matrices. 
For this, the $\chi^2$-function is calculated. In general~\cite{Olive:2016xmw},
\begin{equation} 
\chi^2 (\vec \theta) = \sum_{i = 1}^N 
\frac{\left ( y_i - \mu_i (\vec \theta) \right)^2}{\Delta y_i^2} , 
\label{eq:chi2-def}
\end{equation}
where~$\vec y = \left ( y_1, \ldots, y_N \right )$ is the set of the
experimentally measured values which are assumed to be independent 
and~$\Delta y_i$ are their variances. The quantities $\mu_i (\vec \theta)$ 
are dependent on the unknown parameters which are collected as the vector 
$\vec\theta = \left ( \theta_1, \ldots, \theta_m \right )$ where $m \le N$. 
 For the problem at hand, we take the parameter-dependent  
functions from Eq.~(\ref{Y-states}), $\mu_i (\vec \theta) = M_i$, 
where $i=1,\cdots,4$, and 
\begin{equation}
\vec\theta = \left ( \theta_1,\, \theta_2,\, \theta_3,\, \theta_4 \right ) 
\equiv \left ( M_{00},\, \kappa_{cq},\, a_Y,\, b_Y \right ) .
\label{eq:theta-def-tetra}
\end{equation}

The best-fit estimations of the parameters~$\theta_k$, 
obtained after minimising the $\chi^2$-function, are 
presented in Table~\ref{ali:tbl3}, as central values.   
 Note that each scenario results into two solutions which differ by the sign of the 
best-fit value of~$a_Y$,  in  line with the discussion above.  
The variances of the parameters are also shown in 
Table~\ref{ali:tbl3}, while the correlation matrices are 
collected in Appendix~\ref{app:correlation}. The parameters~(\ref{eq:theta-def-tetra}) are strongly 
correlated as all the correlation moments in the corresponding 
matrices~(\ref{eq:R-matrix-SI-an})--(\ref{eq:R-matrix-SII-ap}) 
are close in magnitude to unity.  To show this, we  plot 
two-dimensional  confidence level (C.L.) contours involving  some of the coefficients.

The correlations among the parameters~$a_Y$ and~$b_Y$ in terms of the~68.3\% 
($\chi^2 = \chi^2_{\rm min} + 2.3$ for two degrees of freedom) and~95.4\% 
($\chi^2 = \chi^2_{\rm min} + 6.18$) C.L. contours are presented 
in Fig.~\ref{fig:correlations-aY-bY}. Similar contours demonstrating correlations 
among~$a_Y$ and~$\kappa_{cq}$ are shown in Fig.~\ref{fig:correlations-aY-kcq}.  
The first and the second rows in these figures correspond to the scenario~I and~II, respectively, 
and in each row, the left panels are plotted for the negative best-fit value of~$a_Y$ 
(c1), while the right panels are for the positive best-fit value (c2). 
Our analysis shows that the scenario I (c1) is not tenable, as, within 95.4\%~C.L. and even 
higher, $a_Y $ remains negative. Thus, the requirement of positive~$a_Y$ disfavours case~1 in the scenario~I. 
In the scenario~II (c1), small positive values of~$a_Y$ are allowed with a relatively 
large probability.  In (c2),  large positive values of~$a_Y$ are predicted for both the scenarios~I and~II.


\section{Discussion} 
\label{discussion}

With the current uncertainty of the experimental scenarios and many parameters one cannot draw 
quantitative conclusions, except observing that the values of the parameters are qualitatively similar to those derived in the $P$-wave $\Omega_c$-states in three of the four solutions. One can, however, underline two criteria that  lead to some preference for the scenario II. 

The first is the value of the chromomagnetic coupling~$\kappa_{cq}$. We  expect the fitted parameter to be close to the analogous parameter derived for the $S$-wave tetraquarks, which is $[\kappa_{cq}]_{S}\simeq 67$~MeV as discussed in~\cite{Maiani:2014aja}. Indeed, there are no reasons to believe that the chromomagnetic coupling~$\kappa_{cq}$ in the diquark should change with the addition of one unit of orbital angular momentum. At 95\% C.L.,
the allowed value of $\kappa_{cq}$ from the $Y$ states in the scenario~II (c1) comes out somewhat smaller
than anticipated, while it is somewhat larger in the scenario~II (c2).  (See, the lower two frames in Fig.~\ref{fig:correlations-aY-kcq}). Thus, this criterion would favour the scenario II.

A second expectation is for the Hamiltonian in Eq.~(\ref{Hamiltonian}) to describe both $S$ and $P$-wave states, with the same value of the diquark mass. As commented in~\cite{Maiani:2014aja}, $Y_2$, which in SII corresponds to $Y(4330)$, is in the same spin state as the $X(3872)$ except that there is a gap in mass between the two, which here is fully accounted by $B_{\mathcal{Q}}$ and  by the spin-orbit interaction.
If this is the case, one can derive the excitation energy of one unit of orbital momentum from the equation
\begin{equation}
M_2 - M[X(3872)] = B_{\mathcal Q} - 2a_Y - [\kappa_{cq}]_P+[\kappa_{cq}]_S.
\label{ve1}
\end{equation}
Using the input from Table~\ref{ali:tbl3} and $[\kappa_{cq}]_S = 67$~MeV, we obtain: 
\bea
&&B_{\mathcal Q} = \left\{ \begin{array}{cc}336~{\rm MeV}, & {\rm SI(c1)}\\545~{\rm MeV}, & {\rm SI(c2)}  \\441~{\rm MeV},& {\rm SII(c1)}  \\596~{\rm MeV},& {\rm SII(c2)}.  \end{array}\right.
\label{ve2}
\eea

We may see what happens if we force the diquark spin-spin couplings to be equal, $[\kappa_{cq}]_P = [\kappa_{cq}]_S = 67$~MeV,  
The $\chi^2$ analysis is redone with four experimental  input values  
and three unknown variables $\theta_k = \left (M_{00}, a_Y, b_Y \right )$. We have 
one degree of freedom and can discriminate minima according to $\chi_{\rm min}^2$. 
The best-fit values and variances of the parameters~$M_{00}$, $a_Y$, and~$b_Y$ (all in MeV) corresponding 
to the minima with the lowest~$\chi_{\rm min}^2$ in each scenario are reported in Table~\ref{chisqd}.
\begin{table}
\begin{center}
\caption{Values of the parameters $M_{00}$, $a_Y$, $b_Y$ (all in MeV), and $\chi_{\rm min}^2$/n.d.f.  
resulting from the $\chi^2$ analysis with fixing $\kappa_{cq} = 67$~MeV.}
\label{chisqd} 
{\normalsize
\begin{tabular}{|c|c|c|c|c|} \hline 
Scenario &      $M_{00}$ &       $a_Y$ &         $b_Y$ & $\chi_{\rm min}^2$/n.d.f. \\ \hline 
      SI & $4321 \pm 79$ &  $2 \pm 41$ & $-141 \pm 63$ & 12.8/1 \\ 
     SII & $4421 \pm  6$ & $22 \pm  3$ & $-136 \pm  6$ &  1.3/1 \\  \hline 
\end{tabular} 
}
\end{center}
\end{table}
There are other minima in both the scenarios, but their $\chi_{\rm min}^2$ are larger, 
and hence we don't discuss the resulting parameters. 

With the parameters from Table~\ref{chisqd} we obtain:

\bea
&&B_{\mathcal Q} = B_{\mathcal Q}(Y)=\left\{ \begin{array}{cc}362~{\rm MeV} & {\rm SI}, \\ 505~{\rm MeV}& {\rm SII}.  \end{array}\right. 
\label{ve3}
\eea
The value for $B_{\mathcal Q} $ from the analysis of the $\Omega_c$-baryon resonances in the diquark model can be obtained from the expression for $M_0$,  given in Eq.~(\ref{eq:M0}). Using the input values of the quark masses and $\kappa_{[ss]}$ from
\cite{Karliner:2016zzc} yields
\be
B_{\mathcal Q}(\Omega_c)= 325~{\rm MeV}.
\label{ve4}
\ee
The values obtained for~$B_{\mathcal Q}$ in Eqs.~(\ref{ve3}) and (\ref{ve4}) can be compared with the orbital angular momentum excitation energy in charmonium, given by the analogous formula
\begin{equation}
B_{\mathcal Q} (c\bar{c})= M(h_c)-\frac{1}{4} \left [ 3 M(J/\psi) + M(\eta_c) \right ] = 457~{\rm MeV} .  
\end{equation}
The combination of $J/\psi$- and $\eta_c$-meson masses eliminates the contribution of  the $S$-wave spin-spin interaction 
in the $J/\psi$-meson, absent in the $h_c$, which has $S_{c\bar c}=0$.

The similarity of the results for different $P$-wave systems with subcomponents in color $3$ and $\bar 3$ is interesting and may suggest more precise calculations for the $Y$ states, with the potential methods applied successfully to charmonia. 
 
 The states with masses~$M_3$ and~$M_4$ are linear combinations of $Y_3$ ($S=0$) and $Y_4$ ($S=2$). 
 We note that in both SI and SII, the eigenvectors corresponding to~$M_3$ and~$M_4$ in~c1 are close to $S=0$ and $S=2$, respectively, 
while in~c2, it is the opposite, {\it i.e.}, they are close to $S=2$ and $S=0$, respectively. 
Table~\ref{ccbarspin} (column 1 and 2) gives the components of the eigenvector associated with~$M_4$, which is called $v_4$, for different
scenarios and solutions. The orthogonal vector $v_3$ is not shown. The eigenvectors carry interesting information; 
 the projection of the eigenvector on the state with $c\bar{c}$ spin =1 is related to the probability of this state to
decay into  a $J/\psi$ ($S_{c\bar{c}}=1$) rather than in $h_c$ ($S_{c\bar{c}}=0$). The fourth column gives the probability of finding $S_{c\bar{c}}=1$ in $v_4$.
The table indicates that $Y(4660)$ in solutions c2 should have a good probability to decay into $h_c$ while in c1 the $J/\psi$
should dominate. This is quantified in the entries in Table~\ref{ccbarspin} (third column).

\begin{table}
 \begin{center}
\caption{First two columns: components of the eigenvector $v_4$ belonging to the highest eigenvalue, $M_4$, in the basis $Y_3,~Y_4$. Third column, Probability of $S_{c\bar c}=1$ in $v_4$.}
{\normalsize
{\begin{tabular}{|c|c|c|c|}
\hline
 & $Y_3,~S=0$  &  $Y_4,~S=2$ &Prob.($S_{c\bar c}$=1) in $v_4$ \\ 
\hline%
SI (c1)& $-0.27$  &  $0.96$ & $0.94$
 \\ 
\hline%
SI (c2) & $0.99$  &  $0.03$ & $0.25$
 \\ 
\hline
SII(c1)& $-0.41$  &  $0.91$ & $0.87$
 \\ 
\hline%
SII (c2) & $-0.99$  &  $0.11$ & $0.26$
 \\ 
\hline
\end{tabular}
}
}
\label{ccbarspin}
\end{center}
\end{table}

\section{The full $L=1$ supermultiplet}  
\label{fullmulti}
We expect many particles in the supermultiplet of $L=1$ diquarkonia, analogous to the  $\chi$-states of charmonia and bottoming.
We find (in~parenthesis~the~multiplicity~of ~the states is given)
\bea
&& 3^{--}~(1);\nonumber \\
&& 2^{--}~(2);~2^{-+}~(2);\nonumber \\
&&1^{--}~(4);~1^{-+}~(2);\nonumber \\
&&0^{--}~(1);~0^{-+}~(2). \nonumber
\label{pwavemult}
\eea
The total number of states coincides with the total number of quark spin and orbital momentum states, i.e., $2^4 \times 3=48$, as one verifies easily. Spin compositions  are given in Tables~\ref{fullmult} and~\ref{ystates},
and  tentative masses are presented inTable~\ref{fullmult}.

Indications exist for two  $0^{-+}$ states. However, in the same channel there should appear two conventional radially excited charmonia, $\eta_c(3S)$ and $\eta_c(4S)$, for a total of four states, with possible mixing and corresponding distortions of the spectrum.

\section{Conclusions}  
\label{conclu}

 We have analysed the masses of the  four lightest $Y$ states, using two experimental scenarios proposed in~\cite{Ali:2017jda,Esposito:2016noz}  and the effective Hamiltonian appropriate for $L=1,~J^{PC}=1^{--}$ tetraquarks, already introduced for the $P$-wave charmonia and for the excited  $\Omega_c$ states.

The current uncertainties on the spectrum of $Y$-states hinder us to reach a completely quantitative conclusion. However, we find  (i) the coefficient of the spin orbit interaction to be positive, within errors, and comparable to that found for the $\Omega_c $ states; (ii) the mass difference of the "bad" and "good" diquarks to be similar to what was found previously for the $S$-wave tetraquarks; (iii) the energy of the orbital excitation is found to be quite comparable to the values for charmonia and  $\Omega_c $; (iv) at variance with  the latter cases, the coefficient of the tensor coupling turns out to be large and negative. The scenario with five $Y$ states proposed in~\cite{Lebed:2016hpi}, including $Y(4008)$ and the two components of the previous $Y(4260)$, is also consistent if one assumes  $Y(4390)$ to be the radial excitation of $Y(4008)$.

Features (i) to (iii) are coherent with our a priori expectations, while we have no particular objection or explanation of (iv).  A slight preference for the Scenario II results, if one insists on enforcing the exact equality of  the mass difference of the "bad" and "good" diquarks in $S$ and $P$ wave states.

Hopefully, some clarification on the composition of the $Y$ states will be provided by BESIII, Belle~II, and LHCb. With precise measurements, the parameters of the effective Hamiltonian can be determined more accurately, providing a quantitative test of the underlying diquark model.
Tetraquarks require many more states in $P$ wave other than the $Y$ states reported in Table~\ref{ystates}, which we have listed in Table~\ref{fullmult}. Tentative mass values are derived from the parameters reported in the second row in
 Table~\ref{chisqd}.

\medskip 
 \begin{table}
\begin{center}
\caption{ Spin composition, couplings and tentative masses (in MeV) of the particles in the $P$-wave supermultiplet, in addition to the states in Table~\ref{ystates}. Mass formulae are derived from Eq.~(\ref{Hamiltonian}), with $M_{00},a_Y,b_Y$ taken from the second line of Table~\ref{chisqd} and $\kappa_{cq}=67$~MeV. $N_1$ is the number of spin-1 "bad" diquarks, defined in
Eq. (\ref{baddqn}).}
\vspace*{2mm}
{\begin{tabular}{|c|c|c|c|c|c|}
\toprule
$J^{PC}$& $| S_{\mathcal{Q}}, S_{\bar{\mathcal{Q}}}; S, L \rangle_J$ &$N_1$& $2{\bm L }{\bm \cdot }{\bm S}$& $S_{12}/4$ & $\begin{array}{c}{\rm Mass (MeV)}\\
{\rm best ~fit ~Table~\ref{chisqd}}\end{array}$\\
\colrule
\hline
$3^{--}$ & $| 1, 1; 2, 1 \rangle_3$ &$2$&4& {\small $-2/5$}&$4630$\\
\hline
$2^{--}$ & $| 1, 1; 2, 1 \rangle_2$&$2$&$-2$&{\small $+7/5$}&$4254$ \\
$2^{--}_a$ & $| \frac{(1, 0)+(0,1)}{\sqrt{2}}; 1, 1 \rangle_2$&$1$&+2& {\small 0}&$4398$ \\
\hline
$2^{-+}$ & $| 1, 1; 1, 1 \rangle_2$&$2$&$+2$& {\small $-1/5$}&$4559$  \\
$2^{-+}_b$ & $| \frac{(1, 0)-(0,1)}{\sqrt{2}}; 1, 1 \rangle_2$&$1$&+2& {\small $0$}&$4398$ \\
\hline
$1^{-+}$ & $| 1, 1; 1, 1 \rangle_1$&$2$&-2& {\small $+1$}&$4308$  \\
$1^{-+}_b$ & $| \frac{(1, 0)-(0,1)}{\sqrt{2}}; 1, 1 \rangle_1$&$1$&-2& {\small $0$}&4310\\
\hline
$0^{-+}$ & $| 1, 1; 1, 1 \rangle_0$&$2$&-4&{\small $-2$}&$4672$ \\
$0^{-+}_b$ & $| \frac{(1, 0)-(0,1)}{\sqrt{2}}; 1, 1 \rangle_0$&$1$&-4& {\small $0$}&4266  \\
\hline
$0^{--}_a$ & $| \frac{(1, 0)+(0,1)}{\sqrt{2}}; 1, 1 \rangle_0$&$1$&-4& {\small $0$}  &4266  \\
\hline
\end{tabular}
}
\label{fullmult}
\end{center}
\end{table}

{\bf \emph{Acknowledgements.}} 
We thank  Marek Karliner, S\"{o}ren Lange,  Richard Lebed, Sheldon Stone, and Changzheng Yuan for helpful discussions.
A.P. acknowledges partial support by the Russian Foundation for Basic Research (Project No. 15-02-06033-a).
One of us (A.A.) would like to thank the CERN Physics Department for the hospitality, where this work originated.
ADP thanks Ryan Mitchell and Alessandro Pilloni for informative comments and discussions.

\appendix



\section{Spin-Orbit,Tensor Coupling and  Wigner's 6-$j$ Symbols}
\label{6jsymbols}

Combining three angular momenta, $j_1,~j_2,~j_3$  to a given $J$, one may follow two paths, characterised by the values of the intermediate angular momentum to which the first two are combined, {\it e.g.} $j_1$ and $j_2$ to $j_{12}$  or $j_2$ and $j_3$ to $j_{23}$, each path corresponding to different base vectors. In the formulae given below, these two bases are characterised as follows
\begin{equation}
 |(j_1,j_2)_{j_{12}},j_3;J\rangle,  \quad\quad |j_1, (j_2,j_3)_{j_{23}};J\rangle,
\end{equation}
or, with a shorter notation
\begin{equation}
 |j_{12},j_3;J\rangle,  \quad\quad |j_1, j_{23};J\rangle,
\end{equation}
where it is understood that $j_1,j_2,j_3$ and $J$ are held fixed. 

Vectors in the two bases are, of course, related by a unitary transformation:
\begin{equation}
|j_1,j_{23};J\rangle=\sum_{j_{12}}C_{j_{23},\,j_{12}}|j_{12},j_3;J\rangle. \label{transf}
\end{equation}
Besides $j_{12}$ and $j_{23}$, the Clebsch--Gordon coefficients $C$ depend upon the angular momenta that are being held fixed, $j_1$, $j_2$, $j_3$ and $J$, that is the $C$s depend on {\it six angular momenta}. To maximise the symmetry properties, one rewrites Eq.~(\ref{transf}) as \cite{edmonds}:
\begin{eqnarray}
|j_1,j_{23};J\rangle &=& \sum_{j_{12}}(-1)^{j_1+j_2+j_3+J}\sqrt{(2j_{12}+1)(2j_{23}+1)}\notag \\
&&\times ~ \begin{Bmatrix}j_1& j_2& j_{12}\\j_3&J&j_{23}\end{Bmatrix}~|j_{12},j_3;J\rangle.  \label{transf6j}
\end{eqnarray}
Wigner's 6-$j$ symbols are represented by the curly brackets. They
 appear in the calculation of the matrix elements of the spin-orbit Hamiltonian or the tensor coupling for two particles with spins $S_1$ and $S_2$ and different masses in the orbital angular momentum $L$. Examples are the $P$-wave $\Omega_c$ baryons and the diquark-antidiquark tetraquarks in $P$-wave, considered in the present paper.

In these cases, to classify states it is convenient  to couple $S_1$ and $S_2$ to a total spin $S$ and couple $S$ to $L$ to obtain the total $J$, that is:
\begin{equation}
\label{lhs}
j_1=L,\quad j_2=S_1,\quad j_3=S_2,\quad  j_{23}=S_1+S_2=S.
\end{equation}
In this basis the matrix elements of the total spin-orbit operator are easily computed according to the formula:
\begin{equation}
\bm L\cdot \bm S=\frac{1}{2}\left [ J(J+1)-L(L +1)-S(S+1) \right ].
\end{equation}

In the spin-orbit interaction and in the tensor coupling, however, one encounters the matrix elements of the operator $\bm L\cdot \bm S_1=\bm j_1\cdot \bm j_2$, which would require a complicated calculation based on writing explicitly the states as products of three angular momentum states and applying the operator $\bm L\cdot \bm S_1$ to them. 

A more convenient way to proceed is  to use Eq.~(\ref{transf6j}) and set 
\begin{equation}
\label{rhs}
j_1=L, \quad j_2=S_1,\quad j_{12}=L+S_1, \quad j_3=S_2.
\end{equation}
In this basis, 
\begin{equation}
\bm L \cdot \bm S_1=\frac{1}{2}\left [ j_{12}(j_{12}+1)-L(L+1)-S_1(S_1+1) \right ] ,
\end{equation}
is diagonal on the basis vectors.

Using Eq.~(\ref{transf6j}), with Eq.~\eqref{lhs} on the lhs and Eq.~\eqref{rhs} on the rhs, one gets  
\begin{eqnarray}
&&\bm L\cdot \bm S_1|S, L;J\rangle \notag \\
&& =\sum_{j_{LS_1}}~(-1)^{L+S_1+S_2+J}\sqrt{(2j_{LS_1}+1)(2S+1)}\notag \\
&& \times   \frac{1}{2}\, [j_{LS_1}(j_{LS_1}+1)-L(L+1)-S_1(S_1+1)]~\notag \\
&&\times \begin{Bmatrix}L& S_1& j_{LS_1}\\S_2&J&S\end{Bmatrix}~|j_{LS_1},S_2;J\rangle.  \label{s1dotl}
\end{eqnarray}
Here, we have used the symbol $j_{12}=j_{LS_1}$, whereas $j_{23}=S$ on the lhs, according to~\eqref{lhs}. It follows that: 
\begin{eqnarray}
&& \langle S^\prime, L; J|\bm L\cdot \bm S_1|S, L;J\rangle=\sqrt{(2S+1)(2S^\prime+1)}\notag \\
&& \times\sum_{j_{LS_1}} \frac{1}{2}\, [j_{LS_1}(j_{LS_1}+1)-L(L+1)-S_1(S_1+1)]\notag \\
&&\times \left ( 2j_{LS_1}+1 \right ) \begin{Bmatrix}L& S_1& j_{LS_1}\\S_2&J&S^\prime\end{Bmatrix} \begin{Bmatrix}L& S_1& j_{LS_1}\\S_2&J&S\end{Bmatrix}, \label{6jcalculation}
\end{eqnarray}
since by definition 
\begin{equation}
\langle j_{12},j_3;J|j_1,j_{23};J\rangle=\langle j_1,j_{23}; J|j_{12},j_3;J\rangle=C_{j_{23}, j_{12}},
\end{equation}
is the coefficient given explicitly in Eq.~\eqref{transf6j}.

Tables of 6-$j$ symbols can be easily implemented  in a computer code and they are already available, making use of the command
 {\tt SixJSymbol[$\{j_1,j_2,j_3\},\{j_4,j_5,j_6\}$]}, in the  symbolic computer algebra system Mathematica~\cite{mathm}.  Therefore the  result in Eq.~(\ref{6jcalculation}) can be obtained with a program of a few lines \cite{thompson}.
In the following, we give the explicit formulae for the cases considered in the paper.

{\bf\emph{ $\Omega_c$ baryons in $P$-wave.}}
The constituents of the states are the $[ss]$-diquark and the charmed quark~$c$ with 
\begin{equation}
j_1=L=1,\quad j_2=S_{[ss]}=1,\quad j_3=S_c=1/2. 
\end{equation}
We will call $j_{12} = j_{LS_{[ss]}}$ and $j_{23} = S = 1/2,\, 3/2$. We have to consider the matrix $\bm L\cdot \bm S_{[ss]}$ in the two cases: $J=1/2$ and $J=3/2$. In the $J=1/2$ case, $j_{LS_{[ss]}}$ can take the values $0,1$ and  Eq.~(\ref{6jcalculation}) reads:
\begin{eqnarray}
&&(\bm L\cdot \bm S_{[ss]})_{J=1/2}\equiv \langle S^\prime, 1;1/2|\bm L\cdot \bm S_{[ss]}|S, 1;1/2\rangle \notag \\
&& = \sqrt{(2S+1)(2S^\prime+1)} \sum_{j_{LS_{[ss]}}=0}^1 \left ( 2 j_{LS_{[ss]}} + 1 \right ) \notag \\
&&\times  \frac{1}{2} \left [ j_{LS_{[ss]}} (j_{LS_{[ss]}} +1) - 4 \right ] \notag \\
&& \times  \begin{Bmatrix}1& 1& j_{LS_{[ss]}}\\1/2&1/2&S^\prime\end{Bmatrix} \begin{Bmatrix}1& 1& j_{LS_{[ss]}}\\1/2&1/2&S\end{Bmatrix}, \label{KR1/2}
\end{eqnarray}
where $S,\, S^\prime=1/2,\, 3/2$. 
This sum can be calculated easily when the required values of 6-$j$ symbols are known: 
\begin{eqnarray} 
&& \begin{Bmatrix}1& 1& 0\\1/2&1/2&1/2\end{Bmatrix} = 
   - \begin{Bmatrix}1& 1& 0\\1/2&1/2&3/2\end{Bmatrix} = \frac{1}{\sqrt 6} , \label{eq:6j-symbols-1} \\ 
&& \begin{Bmatrix}1& 1& 1\\1/2&1/2&1/2\end{Bmatrix} = - \frac{1}{3} , \quad  
   \begin{Bmatrix}1& 1& 1\\1/2&1/2&3/2\end{Bmatrix} = - \frac{1}{6} . \notag 
\end{eqnarray}
Therefore, in the basis of states $({}^4 P_{1/2}, \,  {}^2 P_{1/2})$ we have 
(the notation ${}^{2 S + 1} P_J$ is the same as in~\cite{Karliner:2015ema}):
\begin{equation}
 (\bm L\cdot \bm S_{[ss]})_{J=1/2}= \begin{pmatrix} -5/3 & \sqrt{2}/3\\\sqrt{2}/3 &-4/3 \end{pmatrix}.
\end{equation}
In the same way
\begin{eqnarray}
&&(\bm L\cdot \bm S_{[ss]})_{J=3/2}\equiv \langle S^\prime, 1;3/2|\bm L\cdot \bm S_{[ss]}|S, 1;3/2\rangle \notag \\
&& = \sqrt{(2S+1)(2S^\prime+1)} \sum_{j_{LS_{[ss]}}=1}^2 \left ( 2 j_{LS_{[ss]}} + 1 \right ) \notag \\
&&\times \frac{1}{2} \left [ j_{LS_{[ss]}}(j_{LS_{[ss]}}+1) - 4 \right ] \notag \\
&& \times \begin{Bmatrix}1& 1& j_{LS_{[ss]}}\\1/2&3/2&S^\prime\end{Bmatrix} \begin{Bmatrix}1& 1& j_{LS_{[ss]}}\\1/2&3/2&S\end{Bmatrix},\label{KR3/2}
\end{eqnarray}
with the 6-$j$ symbol values: 
\begin{eqnarray} 
&& \begin{Bmatrix}1& 1& 1\\1/2&3/2&3/2\end{Bmatrix} = \frac{\sqrt{10}}{12} , \quad  
   \begin{Bmatrix}1& 1& 2\\1/2&3/2&1/2\end{Bmatrix} = \frac{1}{2 \sqrt 3} ,  \notag \\   
&& \begin{Bmatrix}1& 1& 2\\1/2&3/2&3/2\end{Bmatrix} = \frac{1}{2 \sqrt{30}} , \label{eq:6j-symbols-2}  
\end{eqnarray}

giving in the basis $({}^4 P_{3/2}, \,  {}^2 P_{3/2})$
\begin{equation}
(\bm L \cdot \bm S_{[ss]})_{J=3/2}= \begin{pmatrix} -2/3 & \sqrt{5}/3\\ \sqrt{5}/3 &2/3  \end{pmatrix}.
\end{equation}
Both results agree with~\cite{Karliner:2017kfm}.

Using the relation $(\bm L\cdot \bm S_c)$ = $(\bm L\cdot \bm S) - (\bm L\cdot \bm S_{[ss]})$, 
it is easy to get the matrices $(\bm L \cdot \bm S_c)_{J=1/2}$ and $(\bm L \cdot \bm S_c)_{J=3/2}$.

{\bf\emph{Diquarkonium in $P$-wave.}}
The constituents are the $[cq]$ diquark and the $[\bar c \bar q]$ antidiquark.
\begin{equation}
\label{assdq}
j_1=L=1,\quad j_2=S_{[cq]}=1,\quad j_3=S_{[\bar c\bar q]}=1.
\end{equation}
Here, $J=1$ and $j_{23} = S = 0,\, 1,\, 2$.  

Note that the state with $S=L=1$ has  positive charge conjugation,~$C$, opposite to the value of~$C$ of the other two states and of the $Y$ states. 

The spin-orbit coupling must be even under~$C$ and, therefore, it is represented by
\begin{equation}
\bm L\cdot(\bm S_{[cq]}+\bm S_{[\bar c \bar q]})=\bm L\cdot \bm S,
\end{equation}
which is diagonal on the states with $S=0,\, 2$.

However, the $C$-even combination of the spin-orbit couplings appearing in the tensor coupling is 
\begin{equation}
(\bm L\cdot \bm S_{[cq]})(\bm L\cdot \bm S_{[\bar c \bar q]})+(\bm L\cdot \bm S_{[\bar c \bar q]})(\bm L\cdot \bm S_{[cq]})
\label{tenspart}
\end{equation}
$\bm L\cdot \bm S_{[cq]}$ is not $C$-invariant and it will mix the states with $S=0,\, 2$ with the other state with $S=1$. The states we have denoted by  $|Y_4\rangle$, $|Y_3\rangle$ have $S=2,~0$ respectively and $C=-1$, and we denote by~$|Y^{(+)}\rangle$ the state with $S=1$ and $C=+1$, see Eq.~(\ref{cplus}). The state~$|Y^{(+)}\rangle$ appears as intermediate state in the products in Eq.~(\ref{tenspart}), giving a contribution to diagonal terms and to non diagonal terms  which mix opposite charge conjugations. The latter, of course, cancel when the sum is taken in Eq.~(\ref{tenspart}).

In conclusion, we have to consider the full $(3\times 3)$ matrix $\bm L\cdot \bm S_{[cq]}$. Using Eqs.~\eqref{6jcalculation} and~\eqref{assdq} we find:
\begin{eqnarray}
&&(\bm L\cdot \bm S_{[cq]})_{J=1}=\langle S^\prime,1; 1|  \bm L\cdot \bm S_{[cq]}|S, 1; 1\rangle \notag \\
&& = \sqrt{(2S+1)(2S^\prime+1)} \sum_{j_{LS_{[cq]}}=0}^2 \left ( 2 j_{LS_{[cq]}} + 1 \right ) \notag \\
&& \times\frac{1}{2}\left [ j_{LS_{[cq]}}(j_{LS_{[cq]}}+1) - 4 \right ] \notag \\  
&& \begin{Bmatrix}1& 1& j_{LS_{[cq]}}\\1&1&S^\prime\end{Bmatrix} \begin{Bmatrix}1& 1& j_{LS_{[cq]}}\\1&1&S\end{Bmatrix}, \label{KR1} 
\end{eqnarray}
where $S,\, S^\prime = 0,\, 1,\, 2$, 
obtaining (for $J=1$): 
\begin{equation}
(\bm L\cdot \bm S_{[cq]})=\begin{pmatrix} -3/2 & 0 & 1/2\sqrt{5/3} \\ 0 & 0 & 2/\sqrt{3}\\ 1/2\sqrt{5/3} & 2/\sqrt{3}&-1/2 \end{pmatrix} , 
\label{mat4q1}
\end{equation}
(states are ordered as  $Y_4,~Y_3,~Y^{(+)}$) in agreement with the result obtained with the direct method of applying the operators ${\bm L}\cdot{\bm S}_{[cq]}$ to the products of angular momentum vectors.
Here, the following values of 6-$j$-symbols are required: 
\begin{eqnarray} 
&& \begin{Bmatrix} 1& 1& 0\\1& 1& 0 \end{Bmatrix} = 
   - \begin{Bmatrix} 1& 1& 0\\1& 1& 1 \end{Bmatrix} = 
   \begin{Bmatrix} 1& 1& 0\\1& 1& 2 \end{Bmatrix} = \frac{1}{3} , \label{eq:6j-symbols-1}  \\ 
&& \begin{Bmatrix} 1& 1& 1\\1& 1& 1 \end{Bmatrix} = 
    \begin{Bmatrix} 1& 1& 1\\1& 1& 2 \end{Bmatrix} = \frac{1}{6} ,  \quad  
   \begin{Bmatrix} 1& 1& 2\\1& 1& 2 \end{Bmatrix} = \frac{1}{30} , \notag 
\end{eqnarray}
and the rest can be obtained with the help of the 6-$j$ symbol symmetry 
under a permutation of columns and interchange of the upper and lower 
arguments in each of any two columns~\cite{edmonds}.  

Using the relation $(\bm L\cdot \bm S_{[\bar c \bar q^\prime]})$=$(\bm L\cdot \bm S)-(\bm L\cdot \bm S_{[cq]})$, 
we also get: 
\begin{equation}
(\bm L\cdot \bm S_{[\bar c \bar q^\prime]})=\begin{pmatrix} -3/2 & 0 & -1/2\sqrt{5/3} \\ 0 & 0 & -2/\sqrt{3}\\ -1/2\sqrt{5/3} & -2/\sqrt{3}&-1/2 \end{pmatrix}, 
\label{mat4q2}
\end{equation}
again in agreement with the result obtained with the direct method.

The states~$2^{- +}$ and~$2^{- -}$ with $J = 2$ are also mixed by the operators  
$(\bm L\cdot \bm S_{[c q]})$ and $(\bm L\cdot \bm S_{[\bar c \bar q^\prime]})$. 
Let us start from the $(2\times 2)$ matrix $(\bm L\cdot \bm S_{[cq]})$ which can 
be obtained as follows: 
\begin{eqnarray}
&&(\bm L\cdot \bm S_{[cq]})_{J=2}=\langle S^\prime, 1; 2|  \bm L\cdot \bm S_{[cq]}|S, 1; 2\rangle \notag \\
&& = \sqrt{(2S+1)(2S^\prime+1)} \sum_{j_{LS_{[cq]}}=1}^2 \left ( 2 j_{LS_{[cq]}} + 1 \right ) \notag \\
&& \times \frac{1}{2}\left [ j_{LS_{[cq]}}(j_{LS_{[cq]}}+1) - 4 \right ] \notag \\  
&& \times \begin{Bmatrix} 1& 1& j_{LS_{[cq]}}\\1&2&S^\prime\end{Bmatrix} \begin{Bmatrix}1& 1& j_{LS_{[cq]}}\\1&2&S\end{Bmatrix}, \label{KR2} 
\end{eqnarray}
where $S,\, S^\prime = 1,\, 2$. 
For the matrices we obtain:  
\begin{equation}
(\bm L \cdot \bm S_{[cq]})_{J=2} = \begin{pmatrix} 1/2 & \sqrt 3/2 \\ \sqrt 3/2 & -1/2 \end{pmatrix} , 
\label{matLQ1}
\end{equation}
and 
\begin{equation}
(\bm L\cdot \bm S_{[\bar c \bar q^\prime]})_{J=2} = \begin{pmatrix} 1/2 & -\sqrt 3/2 \\ -\sqrt 3/2 & -1/2 \end{pmatrix}, 
\label{matLQb1}
\end{equation}
with the states ordered as~$2^{-+}$ and~$2^{--}$.

The averages of the operators $(\bm L\cdot \bm S_{[c q]})$ and $(\bm L\cdot \bm S_{[\bar c \bar q^\prime]})$ 
over the states with $L = 1$ and $J = 0,\, 3$ are required to get the tensor contributions for the masses. 
They can be expressed in terms of the 6-$j$ symbols as follows: 
\begin{equation}
(\bm L \cdot \bm S_{[cq]})_{J=0} = 
(\bm L \cdot \bm S_{[\bar c \bar q^\prime]})_{J=0} = 
-9 \begin{Bmatrix} 1 & 1 & 1 \\ 1 & 0 & 1 \end{Bmatrix}^2 = - 1 , 
\label{matLQ0}  
\end{equation}
\begin{equation}
(\bm L \cdot \bm S_{[cq]})_{J=3} = 
(\bm L \cdot \bm S_{[\bar c \bar q^\prime]})_{J=3} = 
25 \begin{Bmatrix} 1 & 1 & 2 \\ 1 & 3 & 2 \end{Bmatrix}^2 = 1 . 
\label{matLQ3} 
\end{equation}
In addition to the values of the 6-$j$ symbols presented in~(\ref{eq:6j-symbols-1}), 
for the derivation of~(\ref{matLQ1}) and~(\ref{matLQ3}) one needs to have two more: 
\begin{equation} 
\begin{Bmatrix} 1& 1& 1\\1& 2& 2 \end{Bmatrix} = - \frac{1}{2 \sqrt 5} , \quad 
\begin{Bmatrix} 1& 1& 2\\1& 3& 2 \end{Bmatrix} = \frac{1}{5} .     
\label{eq:6j-symbols-4}  
\end{equation}

\section{Correlation Matrices}
\label{app:correlation}  

In this appendix the correlation matrices  in the analysis of the data on $Y$ states are collected. 
We label them in accordance with the notations used in Table~\ref{ali:tbl3}.

SI (c1):  
\begin{equation} 
R = \left ( 
\begin{array}{rrrr}
        1 & -0.890 &  0.995 & -0.990 \\ 
          &      1 & -0.888 &  0.896 \\ 
          &        &      1 & -0.997 \\ 
          &        &        &      1   
\end{array}
\right ).
\label{eq:R-matrix-SI-an}
\end{equation}

SI (c2):   
\begin{equation} 
R = \left ( 
\begin{array}{rrrr}
        1 & -0.927 &  0.974 & -0.960 \\ 
          &      1 & -0.958 &  0.967 \\ 
          &        &      1 & -0.996 \\ 
          &        &        &      1   
\end{array}
\right ).
\label{eq:R-matrix-SI-ap}
\end{equation}

SII (c1):  
\begin{equation} 
R = \left ( 
\begin{array}{rrrr}
        1 &  0.971 &  0.986 & -0.968 \\ 
          &      1 &  0.970 & -0.952 \\ 
          &        &      1 & -0.989 \\ 
          &        &        &      1   
\end{array}
\right ).  
\label{eq:R-matrix-SII-an}
\end{equation}

SII (c2):  
\begin{equation} 
R = \left ( 
\begin{array}{rrrr}
        1 &  0.838 & -0.528 &  0.686 \\ 
          &      1 & -0.534 &  0.674 \\ 
          &        &      1 & -0.972 \\ 
          &        &        &      1   
\end{array}
\right ).
\label{eq:R-matrix-SII-ap}
\end{equation}
%
%
%


\end{document}